\documentclass[tightenlines,eqsecnum,floats,aps,amsmath,amssymb,nofootinbib,
prd,showpacs]{revtex4}
\usepackage{amsmath,amssymb,amsfonts}
\usepackage{graphicx}
\usepackage{enumerate} 
\usepackage{colordvi} 
\usepackage{bm}
\usepackage{epsfig}

\newcommand{\bfv}{\mbox{\boldmath$v$}}

\newcommand{\bfk}{\mbox{\boldmath$k$}}
\newcommand{\bfp}{\mbox{\boldmath$p$}}
\newcommand{\bfq}{\mbox{\boldmath$q$}}

\newcommand{\kk}{\mbox{\boldmath$k$}}
\newcommand{\qq}{\mbox{\boldmath$q$}}
\begin{document}
\title{Non-linear Evolution of Matter Power Spectrum in Modified Theory of
Gravity}
\vfill
\author{Kazuya Koyama$^{1}$, Atsushi Taruya$^{2,3}$, Takashi Hiramatsu$^{4}$}
\bigskip
\address{$^1$Institute of Cosmology \& Gravitation, University of Portsmouth,
Portsmouth, Hampshire, PO1 2EG, UK}
\address{$^2$Research Center for the Early Universe, School of Science,
University of Tokyo, Bunkyo-ku, Tokyo 113-0033, Japan}
\address{$^3$Institute for the Physics and Mathematics of the Universe,
University of Tokyo, Kashiwa, Chiba 277-8568, Japan}
\address{$^4$ Institute for Cosmic Ray Research, University of Tokyo,
Kashiwa, Chiba 277-8582, Japan}
\bigskip
\vfill
\date{today}

\begin{abstract}
We present a formalism to calculate the non-linear matter power spectrum in
modified gravity models that explain the late-time acceleration
of the Universe without dark energy.
Any successful modified gravity models should contain a mechanism
to recover General Relativity (GR) on small scales in order to
avoid the stringent constrains on deviations from GR at solar
system scales. Based on our formalism,
the quasi non-linear power spectrum in the
Dvali-Gabadadze-Porratti (DGP)
braneworld models and  $f(R)$ gravity models are derived
by taking into account the mechanism to recover GR properly.
We also extrapolate our predictions to fully non-linear
scales using the Parametrized Post Friedmann (PPF) framework.
In DGP and $f(R)$ gravity models,
the predicted non-linear power spectrum is shown to reproduce N-body results.
We find that the mechanism to recover GR suppresses the
difference between the modified gravity models and dark energy models
with the same expansion history, but the difference remains large at
weakly non-linear regime in these models. Our formalism is applicable to a wide
variety of modified gravity models and it is ready to use once
consistent models for modified gravity are developed.
\end{abstract}
\pacs{98.80.-k}
\maketitle

\section{Introduction}
\label{sec:intro}
The late-time acceleration of the Universe is surely the most
challenging problem in cosmology. Within the framework of general
relativity (GR), the acceleration originates from dark energy.
The simplest option is the cosmological constant. However, in order to
explain the current acceleration of the Universe, the required value of
the cosmological constant must be incredibly small. Alternatively, there
could be no dark energy, but a large distance modification of GR may
account for the late-time acceleration of the Universe. Recently considerable
efforts have been made to construct models for modified gravity as an
alternative
to dark energy and distinguish them from dark energy models by observations
(see \cite{Nojiri:2006ri, Koyama:2007rx, Durrer:2008in, Durrer:2007re} for
reviews).

Although fully consistent models have not been constructed yet,
some indications of the nature of the modified gravity models
have been obtained. In general, there are three regimes of gravity
in modified gravity models \cite{Koyama:2007rx, Hu07}.
On the largest scales, gravity must be modified
significantly in order to explain the late time acceleration without
introducing dark energy. On the smallest scales, the theory must approach GR
because there exist stringent constraints on the deviation from GR at solar
system scales. On intermediate scales between the cosmological horizon
scales and the solar system scales, there can be still a deviation
from GR. In fact, it is a very common feature in modified gravity models
that there is a significant deviation from GR on large scale structure
scales. This is due to the fact that, once we modify GR, there arises a new
scalar degree of freedom in gravity. This scalar mode changes gravity even
below the length scale where the modification of the tensor sector of
gravity becomes significant, which causes the cosmic acceleration.

Therefore, large scale structure of the Universe offers the best opportunity
to distinguish between modified gravity models and dark energy models in GR
\cite{Uzan:2000mz, Lue:2003ky, ishak05, Knox:2006fh, koyama06, Chiba:2007rb,
Amendola:2007rr, Yamamoto:2007gd, Yamamoto:2008gr, Song:2008qt, Kunz:2006ca,
Jain:2007yk, Song:2008vm, Song:2008xd, Afshordi:2008rd, Zhao:2008bn, Zhao:2009fn}.
The expansion history of the Universe determined by the Friedman equation can be
completely the same in modified gravity models and dark energy models. In fact,
it is always possible to find a dark energy model that can mimic
the expansion history of the Universe in a given modified gravity model by
tuning the equation of state of dark energy. However, this degeneracy
can be broken by the growth rate of structure formation. This is because the
scalar degree of freedom in modified gravity models changes the strength of
gravity on sub-horizon scales and thus changes the growth rate of structure
formation. Thus combining the geometrical test and structure formation test,
one can distinguish between dark energy models and modified gravity models.

However, there is a subtlety in testing modified gravity models using large
scale structure of the Universe. In any successful modified gravity models,
we should
recover GR on small scales. Indeed, unless there is an additional mechanism
to screen the scalar interaction which changes the growth rate of structure
formation, the modification of gravity contradicts to the stringent constraints
on the deviation from GR at solar system scales. This mechanism affects
the non-linear clustering of dark matter. We expect that the power-spectrum
of dark matter perturbations approaches the one in the GR dark energy model
with the same expansion history of the Universe because the modification
of gravity disappears on small scales.
Then the difference between a modified gravity model and a dark energy model
with the same expansion history becomes smaller on smaller scales. This
recovery of GR has important implications for weak lensing measurements
because the strongest signals in weak lensing measurements come from
non-linear scales. We should note that the non-linear power spectrum
is also sensitive to the properties of dark energy \cite{Alimi:2009zk}.

In almost all of the literature, the non-linear power spectrum in
modified gravity models was derived using the mapping formula between the linear
power spectrum and the non-linear power spectrum. This is equivalent to assume
that gravity is modified down to small scales in the same way as in the
linear regime which contradicts to the solar system constraints. Thus this
approach overestimates the difference between modified gravity models and dark
energy models. This was explicitly shown by N-body simulations in the context
of $f(R)$ gravity \cite{Oyaizu:2008sr, Oyaizu:2008tb, Schmidt:2008tn}.
In $f(R)$ gravity, the Einstein-Hilbert action is replaced by
an arbitrary function of Ricci curvature (see \cite{Sotiriou:2008rp, Capozziello:2007ec}
for a review).
This model is equivalent with the Brans-Dicke (BD) theory with non-trivial
potential \cite{Gottlober:1989ww, Wands:1993uu, Magnano:1993bd}.
The BD scalar mediates an additional gravitational interaction
that enhances the gravitational force below the Compton wavelength of
the BD scalar. If the mass of the BD scalar becomes larger in a dense
environment like in the solar system, the Compton wavelength becomes
short and we can recover GR. This is known as the chameleon mechanism
\cite{Khoury:2003rn}. In the context of $f(R)$ gravity, by tuning the function $f$,
it is possible to make the Compton wavelength of the BD scalar short at solar system
scales and screen the BD scalar interaction
\cite{Hu:2007nk, Starobinsky:2007hu, Appleby:2007vb, Cognola:2007zu}.
N-body simulations show that, due to this mechanism, the deviation of the
non-linear power spectrum from GR is suppressed on small scales
\cite{Oyaizu:2008sr, Oyaizu:2008tb, Schmidt:2008tn}. It was shown
that the mapping formula failed to describe this recovery of GR and it
overestimated the deviation from GR.

In this paper, we develop a formalism to treat the quasi non-linear evolution of
the power spectrum in modified gravity models by properly taking into account
the mechanism to recover GR on small scales. Our formalism is based on
the closure approximation which gives a closed set of evolution equations
for the matter power spectrum \cite{Taruya:2007xy}.
These evolution equations reproduce the
one-loop results of the standard perturbation theory (SPT) by replacing the
quantities in the non-linear terms with linear-order ones.
The SPT in GR is tested against N-body simulations extensively
recently and it has been shown that, at the quasi non-linear regime, it can
predict the power-spectrum with a sub-percent accuracy \cite{Nishimichi:2008ry}.
Although the validity regime of the perturbation theory is limited, it is the most
relevant regime to distinguish between modified gravity models and dark energy
models in GR because the difference in the two models is large
in the linear and quasi-non-linear regime. We developed a general
formalism which can be applied to many modified gravity models
including well studied models such as $f(R)$ gravity
and braneworld models.

This paper is organized as follows. In section II, we introduce
effective equation for quasi-static perturbations to describe the Newtonian limit of
gravity. Basic equations which are necessary to compute the power
spectrum are presented in section III.
In section IV, the non-linear evolution equations of the
power spectrum are derived based on the closure approximation proposed
by Ref.~\cite{Taruya:2007xy}.
The closure approximation is one of the non-perturbative prescriptions for
comupting non-linear power spectrum, and it is shown to be equivalent to
the one-loop level of renormalized perturbation theory by Crocce and
Scoccimarro \cite{Crocce:2005xy} and the 2PI effective action method
by Valageas \cite{Valageas:2006bi}.
By replacing all the quantities in non-linear terms with linear-order
ones, the so-called one-loop power spectrum
is obtained numerically which describes the leading-order non-linear
corrections.
In section V, we apply this formalism to Dvali-Gabadadze-Porratti
(DGP) braneworld models. In the case of DGP models, we can derive the quasi
non-linear power spectrum analytically. We test our numerical code
to solve the closure equation against the analytical results.
The quasi non-linear spectrum is derived also in $f(R)$ gravity models.
In this case, our results are compared with N-body simulations.
In section VI, we apply the Parametrized post-Friedman (PPF) framework
to predict the fully non-linear power spectrum. Using the solutions in
the perturbation theory, we determine a parameter in the PPF framework.
Then we predict the non-linear power spectrum by extrapolating this
parameter to fully non-linear scales.
The predictions of the PPF formalism are compared with N-body simulations.
Section VII is devoted to conclusions. In appendix A, a numerical scheme to
solve the closure equations is presented. In appendix B, we derive
the quasi non-linear power spectrum in DGP models analytically.

\section{Quasi-static perturbations in modified gravity models}
\label{sec:quasi-static_PT}

We consider perturbations around the Friedman-Robertson-Walker
universe described in the Newtonian gauge:
\begin{equation}
ds^2=-(1+2\psi)dt^2+a^2(1+2\phi)\delta_{ij}dx^idx^j.
\end{equation}
We will study the evolution of matter fluctuations inside the Hubble horizon.
Then we can use the quasi-static approximation and neglect the time derivatives
of the perturbed quantities compared with the spatial derivatives. As mentioned
in the introduction, the large distance modification of gravity, which is
necessary to explain the late-time acceleration, generally modifies gravity
even on sub-horizon scales due to the introduction of a new scalar degree
of freedom. This modification of gravity due to the scalar mode can be
described by Brans-Dicke (BD) gravity. Under the quasi-static approximations,
perturbed modified Einstein equations give
\begin{eqnarray}
\phi + \psi &=& - \varphi,
\label{eq1}
\\
\frac{1}{a^2} \nabla^2 \psi &=& 4 \pi G \rho_m \delta
-\frac{1}{2 a^2} \nabla^2 \varphi,
\label{eq:Poisson_eq}
\\
(3 +2 \omega_{\rm BD}) \frac{1}{a^2} \nabla^2 \varphi
&=&- 8 \pi G \rho_m \delta,
\end{eqnarray}
where $G $ is the Newton constant measured in Cavendish-like experiments,
$\rho_m$ is the background dark matter energy density and $\delta$ is
dark matter density perturbations. Under the quasi-static approximations
$\omega_{\rm BD}$ can be any function of time.
In general, modified gravity models that
explain the late time acceleration predict $\omega_{\rm BD} \sim O(1)$ on
sub-horizon scales today.
This would contradict to the solar system constraints which require
$\omega_{\rm BD} > 40000$. However, this constraint can be applied only
when the BD scalar has no potential and no self-interactions. Thus, in
order to avoid this constraint, the BD scalar should acquire some
interaction terms on small scales.
In general we expect that the BD scalar field equation is given by
\begin{equation}
(3 +2 \omega_{\rm BD}) \frac{1}{a^2} k^2 \varphi
= 8 \pi G \rho_m \delta - {\cal I}(\varphi),
\label{eq:BD_eq}
\end{equation}
in a Fourier space. Here the interaction term ${\cal I}$ can be expanded as
\begin{eqnarray}
{\cal I} (\varphi)
&=& M_1(k) \varphi + \frac{1}{2} \int \frac{d^3 \bfk_1 d^3 \bfk_2}
{(2 \pi)^3} \delta_D(\bfk -\bfk_{12}) M_2(\bfk_1, \bfk_2)
\varphi(\bfk_1) \varphi(\bfk_2) \nonumber\\
&&+ \frac{1}{6}
\int \frac{d^3 \bfk_1 d^3 \bfk_2 d^3 \bfk_3}{(2 \pi)^6}
\delta_D(\bfk - \bfk_{123}) M_3(\bfk_1, \bfk_2, \bfk_3)
\varphi(\bfk_1)\varphi(\bfk_2) \varphi(\bfk_3) + ...
\end{eqnarray}
where $\bfk_{ij}=\bfk_i+\bfk_j$ and $\bfk_{ijk}=\bfk_i+\bfk_j+\bfk_k$.

We should emphasize that effective equations (\ref{eq1}), (\ref{eq:Poisson_eq})
and (\ref{eq:BD_eq}) in the BD theory can be applied only to quasi-static
perturbations. We allow the time dependence of the BD parameter in these
effective equations but this does not necessarily mean that we are considering
scalar tensor theory where the BD parameter is a function of the BD scalar.
The effective equations are applicable to scalar tensor theory by adding
appropriate non-linear interaction terms in ${\cal I}$. Although this is an
interesting possibility (see \cite{Nagata:2002tm, EspositoFarese:2000ij, Riazuelo:2001mg}
for the analysis of perturbations in scalar tensor theory), we will not consider this
possibility in this paper. We also note that a general parametrization of linear perturbations
in modified gravity was developed in \cite{Bertschinger:2008zb} and
a similar parametrisation of quasi-static perturbations to ours was considered
at linearized level in \cite{Zhao:2008bn}.

In this paper, we consider two known mechanisms where the non-linear interaction terms
${\cal I}$ are responsible for the recovery of GR on small scales.
There are other possibilities to recover GR on small scales for example
by decoupling baryons, but we focus on the following two possibilities
in this paper. One is the chameleon mechanism \cite{Khoury:2003rn}. In this case,
the BD scalar has a non-trivial potential, and acquires a mass.
Then, the BD scalar mediates the Yukawa-type
force and the interaction decays exponentially above
the length scale determined by the inverse of the mass, the
Compton wavelength. Because of this, the scalar interaction is hidden
above the Compton wavelength, and GR is recovered.
The BD scalar is coupled to the trace of the energy
momentum tensor. Thus the effective potential depends on the
energy density of the environment.
The potential is tuned so that the mass of the BD scalar becomes
large for a dense environment such as the solar system.
In order for the chameleon mechanism to work, the scalar fields
needs a runaway potential to be efficient \cite{Khoury:2003rn}.
Then the Compton wavelength becomes very short for a dense environment
and the scalar mode is effectively hidden. In this paper,
we deal with this mechanism perturbatively. $M_1$ determines the mass
term in the cosmological background. The higher
order terms $M_i, (i >1)$ describe the change of the mass
term due to the change of the energy density. If the
chameleon mechanism is at work, the effective mass becomes
larger when the density fluctuations become non-linear.

The other mechanism relies on the existence of the non-linear
derivative interactions. A typical example is the Dvali-Gabadadze-Porratti
(DGP) braneworld model where we are supposed to be living on a 4D brane
in a 5D Minkowski spacetime \cite{dvali00}.
In this model, the BD scalar is identified as
the brane bending mode which describes the deformation of the
4D brane in the 5D bulk spacetime. The brane bending mode has
a large second-order term in the equation of motion which cannot be
neglected even when the metric perturbations remain linear.
This corresponds to the existence of a large $M_2(k)$ term
\cite{Luty:2003vm, Tanaka:2003zb, Koyama:2007ih}.
It has been shown
that once this second order term dominates over the linear term, the scalar
mode is hidden and the solutions for metric perturbations approach GR solutions.
For a static spherically symmetric source, we can identify the length
scale below which the second order interaction becomes important.
This length scale is known as the Vainshtein radius \cite{Deffayet:2001uk}.
In the cosmological
situation, it is expected that once the density perturbations become non-linear,
the second order term becomes important and we recover GR. In the next section,
we apply the perturbation theory to solve the equations (\ref{eq1}), (\ref{eq:Poisson_eq})
and (\ref{eq:BD_eq}). Thus we
only keep up to the third order in the expansion of ${\cal I}$
which is necessary to calculate the quasi non-linear power spectrum.

The evolution equations for matter perturbations are obtained
from the conservation of energy momentum tensor, the continuity equation
and the Euler equation:
\begin{eqnarray}
&&\frac{\partial \delta}{\partial t}+\frac{1}{a}\nabla\cdot
[(1+\delta)\bfv]=0,
\label{eq:continuity_eq}\\
&&\frac{\partial \bfv}{\partial t}+H\bfv+\frac{1}{a}(\bfv\cdot\nabla)\bfv=
-\frac{1}{a}\nabla\psi.
\label{eq:Euler_eq}
\end{eqnarray}
Eqs.~(\ref{eq:Poisson_eq}), (\ref{eq:BD_eq}), (\ref{eq:continuity_eq}) and
(\ref{eq:Euler_eq}) are the basic equations that have to be
solved. In the next section, we derive evolution equation for
perturbations in a compact form.

\section{Evolution equations for perturbations}
\label{sec:closure_approx}
Assuming the irrotationality of fluid quantities, the velocity field
$\bfv$ is expressed in terms of velocity divergence
$\theta\equiv\nabla\cdot\bfv/(a\,H)$. Then the Fourier transform of the
fluid equations (\ref{eq:continuity_eq}) and (\ref{eq:Euler_eq}) become
\begin{eqnarray}
&&H^{-1}\frac{\partial \delta(\bfk)}{\partial t}+\theta(\bfk) =-
\int\frac{d^3\bfk_1d^3\bfk_2}{(2\pi)^3}\delta_{\rm D}(\bfk-\bfk_1-\bfk_2)
\alpha(\bfk_1,\bfk_2)\,\theta(\bfk_1)\delta(\bfk_2),
\label{eq:Perturb1}\\
&&H^{-1}\frac{\partial \theta(\bfk)}{\partial t}+
\left(2+\frac{\dot{H}}{H^2}\right)\theta(\bfk)
-\left(\frac{k}{a\,H}\right)^2\,\psi(\bfk)=
-\frac{1}{2}\int\frac{d^3\bfk_1d^3\bfk_2}{(2\pi)^3}
\delta_{\rm D}(\bfk-\bfk_1-\bfk_2)
\beta(\bfk_1,\bfk_2)\,\theta(\bfk_1)\theta(\bfk_2),
\label{eq:Perturb2}
\end{eqnarray}
where the kernels in the Fourier integrals, $\alpha$  and
$\beta$, are given by
\begin{eqnarray}
\alpha(\bfk_1,\bfk_2)=1+\frac{\bfk_1\cdot\bfk_2}{|\bfk_1|^2},
\quad\quad
\beta(\bfk_1,\bfk_2)=
\frac{(\bfk_1\cdot\bfk_2)\left|\bfk_1+\bfk_2\right|^2}{|\bfk_1|^2|\bfk_2|^2}.
\end{eqnarray}
As for the Poisson equation (\ref{eq:Poisson_eq}),
the potential $\psi$ is couples to $\delta$ through the BD scalar $\varphi$
in a fully non-linear way due to the interaction term ${\cal I}$.
To derive closed equations for $\delta$ and $\theta$, we must employ the
perturbative approach to Eq.~(\ref{eq:BD_eq}). By solving Eq.~(\ref{eq:BD_eq})
perturbatively assuming $\varphi<1 $, $\psi$ can be expressed in
terms of $\delta$ as
\begin{equation}
-\left(\frac{k}{a}\right)^2\psi=\frac{1}{2}\,\kappa^2 \,\rho_{m}\,
\left[1+ \frac{1}{3}
\frac{(k/a)^2}{\Pi(k)}\right]
\,\delta(\bfk) +\frac{1}{2}\left(\frac{k}{a}\right)^2\,S(\bfk),
\label{eq:modified_Poisson}
\end{equation}
where
\begin{equation}
\Pi(k) = \frac{1}{3} \left((3+2 \omega_{\rm BD}) \frac{k^2}{a^2} + M_1 \right),
\end{equation}
and $\kappa^2 = 8 \pi G$.
The function $S(\bfk)$ is the non-linear source term which is
obtained perturbatively using (\ref{eq:Poisson_eq}) as
\begin{eqnarray}
S(\bfk)&=&-\frac{1}{6\Pi(\bfk)}\,
\left(\frac{\kappa^2\,\rho_{m}}{3}\right)^2
\int\frac{d^3\bfk_1d^3\bfk_2}{(2\pi)^3}\,
\delta_{\rm D}(\bfk-\bfk_{12}) M_2(\bfk_1, \bfk_2)
\frac{\delta(\bfk_1)\,\delta(\bfk_2)}{\Pi(\bfk_1)\Pi(\bfk_2)}
\nonumber\\
&&-\frac{1}{18\,\Pi(\bfk)}\,
\left(\frac{\kappa^2\,\rho_{m}}{3}\right)^3\,
\int\frac{d^3\bfk_1d^3\bfk_2d^3\bfk_3}{(2\pi)^6}
\delta_{\rm D}(\bfk-\bfk_{123})
\left\{M_3(\bfk_1, \bfk_2, \bfk_3)-
\frac{M_2(\bfk_1,\bfk_2+\bfk_3) M_2(\bfk_2, \bfk_3)}{\Pi(\bfk_{23})}\right\}
\nonumber\\
&& \times \frac{\delta(\bfk_1)\,\delta(\bfk_2)\delta(\bfk_3)}
{\Pi(\bfk_1)\Pi(\bfk_2)\Pi(\bfk_3)}. \label{eq:Perturb3}
\end{eqnarray}
The expression (\ref{eq:Perturb3}) is valid up to the third-order in $\delta$.

The perturbation equations
(\ref{eq:Perturb1}), (\ref{eq:Perturb2}) and (\ref{eq:modified_Poisson})
can be further reduced to a compact form by introducing the
following quantity:
\begin{equation}
\Phi_a(\bfk)=\left(
\begin{array}{c}
\delta(\bfk) \\
-\theta(\bfk)
\end{array}
\right).
\end{equation}
We can write down the basic equations in a single form as
\begin{eqnarray}
&&\frac{\partial \Phi_a(\bfk;\tau)}{\partial\tau} +
\Omega_{ab}(k;\tau)\,\Phi_b(\bfk;\tau) =
\int\frac{d^3\bfk_1d^3\bfk_2}{(2\pi)^3}\,\delta_{\rm D}(\bfk-\bfk_{12})\,
\gamma_{abc}(\bfk_1,\bfk_2;\tau)\,\Phi_b(\bfk_1;\tau)\Phi_c(\bfk_2;\tau)
\nonumber\\
&&\quad\quad\quad\quad\quad\quad\quad\quad\quad\quad
+\int \frac{d^3\bfk_1d^3\bfk_2d^3\bfk_3}{(2\pi)^6}\,
\delta_{\rm D}(\bfk-\bfk_{123})\,
\sigma_{abcd}(\bfk_1,\bfk_2,\bfk_3;\tau)\,
\Phi_b(\bfk_1;\tau)\Phi_c(\bfk_2\tau)\Phi_d(\bfk_3;\tau),
\label{eq:basic_eq}
\end{eqnarray}
where the time variable $\tau$ is defined by $\tau=\ln a(t)$.
The matrix $\Omega_{ab}$ is given by
\begin{equation}
\Omega_{ab}(k;\tau)=\left(
\begin{array}{cc}
{\displaystyle 0} & \,\,{\displaystyle -1 }\\
\\
{\displaystyle -\frac{\kappa^2}{2}\frac{\rho_{m}}{H^2}
\left[1+
\frac{1}{3}\frac{(k/a)^2}{\Pi(k)}\right] }&
\,\,{\displaystyle 2+\frac{\dot{H}}{H^2}}
\end{array}
\right).
\end{equation}
From the $(2,1)$ component of $\Omega_{ab}$, we can define the effective
Newton constant as
\begin{equation}
G_{\rm eff} =  G \left[1 + \frac{1}{3} \frac{(k/a)^2}{\Pi(k)}
\right].
\label{eq:effective_G}
\end{equation}
If $M_1$=0, the effective Newton constant is given by
\begin{equation}
G_{\rm eff} = \frac{2 (2+\omega_{\rm BD})}{3 +2 \omega_{\rm BD} }G.
\label{eq:effectiveG}
\end{equation}
See Ref.~\cite{Fujii:2003pa} for a review on the BD theory.
For a positive $\omega_{\rm BD} >0$, the effective gravitational
constant is larger than GR and the gravitational force is
enhanced. On the other hand, if $M_1 \gg k^2/a^2$, $G_{\rm eff}$ becomes
$G$.
The quantity $\gamma_{abc}$ is the vertex function as in the GR case,
but new non-vanishing components arise in the case of modified gravity:
\begin{eqnarray}
\gamma_{abc}(\bfk_1,\bfk_2;\tau)=
\left\{
\begin{array}{lcl}
{\displaystyle \frac{1}{2}\,\alpha(\bfk_2,\bfk_1) }&;& (a,b,c)=(1,1,2), \\
\\
{\displaystyle \frac{1}{2}\,\alpha(\bfk_1,\bfk_2) }&;& (a,b,c)=(1,2,1), \\
\\
{\displaystyle -\frac{1}{12H^2}\,
\left(\frac{\kappa^2\,\rho_{m}}{3}\right)^2 \left(
\frac{k_{12}^2}{a^2} \right)
\frac{M_2(\bfk_1, \bfk_2)}{\Pi(\bfk_{12})\Pi(\bfk_1)\Pi(\bfk_2)} }&;& (a,b,c)=(2,1,1), \\
\\
{\displaystyle \frac{1}{2}\,\beta(\bfk_1,\bfk_2) }&;& (a,b,c)=(2,2,2), \\
\\
{\displaystyle 0 } &;& \mbox{otherwise}.
\end{array}
\right.
\end{eqnarray}
Here $\gamma_{211}$ is absent in GR. Note that the symmetric properties
of the vertex function, $\gamma_{abc}(\bfk_1,\bfk_2; \tau)=\gamma_{acb}(\bfk_2,\bfk_1; \tau)$,
still hold in the modified theory of gravity.
In Eq.~(\ref{eq:basic_eq}), there appears another vertex
function coming from the non-linearity of Poisson equation. The
explicit form of the higher-order vertex function $\sigma_{abcd}$ is given by
\begin{eqnarray}
\sigma_{abcd}(\bfk_1,\bfk_2,\bfk_3;\tau)=
\left\{
\begin{array}{lcl}
{\displaystyle -\frac{1}{36 H^2}\,
\left(\frac{\kappa^2 \rho_{m}}{3}\right)^3
\left(\frac{k_{123}^2}{a^2}\right)
\frac{M_3(\bfk_1, \bfk_2, \bfk_3)}
{\Pi(\bfk_{123})\Pi(\bfk_1)\Pi(\bfk_2)\Pi(\bfk_3)} }&& \\
{\displaystyle
\times\left[1-\frac{1}{3}\frac{1}{M_3(\bfk_1, \bfk_2, \bfk_3)}
\left\{\frac{M_2(\bfk_1, \bfk_2+\bfk_3)M_2(\bfk_2, \bfk_3)}{\Pi(\bfk_{23})}+
\mbox{perm.}\right\}\right] } &;& (a,b,c,d)=(2,1,1,1),
\\
{\displaystyle 0 } &;& \mbox{otherwise}.
\end{array}
\right.
\end{eqnarray}
Again this term is absent in GR. The vertex function $\sigma_{abcd}(\bfk_1,\bfk_2,\bfk_3;\tau)$
defined above is invariant
under the permutation of $b\leftrightarrow c \leftrightarrow d$ or
$\bfk_1\leftrightarrow \bfk_2 \leftrightarrow \bfk_3$.

\section{Evolution equations for the power spectrum}

In this paper, we are especially interested in the evolution of
the matter power spectrum defined by
\begin{equation}
\Bigl\langle\Phi_a(\bfk;\tau)\Phi_b(\bfk';\tau)\Bigr\rangle =
(2\pi)^3\,\delta_{\rm D}(\bfk+\bfk')\,P_{ab}(|\bfk|;\tau).
\label{eq:def_pk}
\end{equation}
Here the bracket $\langle \cdot \rangle$ stands for the ensemble average.
Note that we obtain the three different power spectra: $P_{\delta\delta}$
from $(a,b)=(1,1)$, $-P_{\delta\theta}$ from $(a,b)=(1,2)$ and $(2,1)$,
and $P_{\theta\theta}$ from $(a,b)=(2,2)$.

Let us consider how to compute the power spectrum.
In the standard perturbation theory (SPT), we first
solve Eq.(\ref{eq:basic_eq}) by expanding the quantity $\Phi_a$ as
$\Phi_a=\Phi_a^{(1)}+\Phi_a^{(2)}+\cdots$. Substituting
the perturbative solutions into the definition (\ref{eq:def_pk}),
we obtain the weakly non-linear corrections to
the power spectrum. This treatment is straightforward, but it is not
suited for numerical calculations.
Furthermore, successive higher-order corrections
generally converge poorly and SPT will be
soon inapplicable at the late-time stage of the non-linear evolution.
Here in order to deal with modified gravity models,
in which analytical calculation is intractable in many cases,
we take an alternative approach. Our approach is based
on the closure approximation proposed by Ref.~\cite{Taruya:2007xy},
by which the evolution of the power spectrum is obtained numerically
by solving a closed set of evolution equations.

Provided the basic equation (\ref{eq:basic_eq}),
evolution equations for the power
spectrum can be derived by truncating an infinite chain
of the moment equations with a help of perturvative calculations
called the closure approximation.
We skip the details of the derivation and present the final results.
Readers who are interested in the derivation can refer to
Ref.~\cite{Taruya:2007xy}.
The resultant evolution equations are the coupled equations
characterized by the three statistical quantities including the power
spectrum. We define
\begin{eqnarray}
&&\Bigl\langle\Phi_a(\bfk;\tau)\Phi_b(\bfk';\tau')\Bigr\rangle =
(2\pi)^3\,\delta_{\rm D}(\bfk+\bfk')\,R_{ab}(|\bfk|;\tau,\tau')
\quad; \quad\quad \tau>\tau',
\nonumber\\
&&\Bigl\langle\frac{\delta \Phi_a(\bfk;\tau)}{\delta \Phi_b(\bfk';\tau')}
\Bigr\rangle = \delta_{\rm D}(\bfk-\bfk')\,G_{ab}(\bfk|\tau,\tau')
\quad; \quad\quad \tau\geq\tau'.
\end{eqnarray}
The quantities $R_{ab}$ and $G_{ab}$  denote the cross
spectra between different times and the non-linear propagator, respectively.
Note that $R_{ab}\ne R_{ba}$, in general.
Then, the closure equations become
\begin{eqnarray}
\widehat{\Sigma}_{abcd}(k;\tau)\,P_{cd}(k;\tau)
&=&\int\frac{d^3\bfq}{(2\pi)^3}\Bigl[
\gamma_{apq}(\bfq,\bfk-\bfq;\tau)\,F_{bpq}(-\bfk,\bfq,\bfk-\bfq;\tau)
+\gamma_{bpq}(\bfq,-\bfk-\bfq;\tau)\,F_{apq}(\bfk,\bfq,-\bfk-\bfq;\tau)
\Bigr]
\nonumber\\
&&+3 \int\frac{d^3\bfq}{(2\pi)^3}\Bigl[
\sigma_{apqr}(\bfq,-\bfq,\bfk;\tau)P_{pq}(q;\tau)P_{rb}(k;\tau)
+\sigma_{bpqr}(\bfq,-\bfq,-\bfk;\tau)\,P_{pq}(q;\tau)P_{ra}(k;\tau)
\Bigr],\nonumber\\
\label{eq:CLA_eq1} \\
\widehat{\Lambda}_{ab}(k;\tau)\,R_{bc}(k;\tau,\tau')
&=&\int\frac{d^3\bfq}{(2\pi)^3}
\gamma_{apq}(\bfq,\bfk-\bfq;\tau)\,K_{cpq}(-\bfk,\bfq,\bfk-\bfq;\tau,\tau')
\nonumber\\
&&+3 \int\frac{d^3\bfq}{(2\pi)^3}
\sigma_{apqr}(\bfq,-\bfq,\bfk;\tau)P_{pq}(q;\tau)R_{rc}(k;\tau,\tau'),
\label{eq:CLA_eq2}\\
\widehat{\Lambda}_{ab}(k;\tau)\,G_{bc}(k|\tau,\tau')
&=&4\int_{\tau'}^{\tau}d\tau'' \int\frac{d^3\bfq}{(2\pi)^3}
\gamma_{apq}(\bfq,\bfk-\bfq;\tau)\,\gamma_{lrs}(-\bfq,\bfk;\tau'')
G_{ql}(|\bfk-\bfq||\tau,\tau'')R_{pr}(q;\tau,\tau'')G_{sc}(k|\tau'',\tau'),
\nonumber\\
&&+3 \int\frac{d^3\bfq}{(2\pi)^3}
\sigma_{apqr}(\bfq,-\bfq,\bfk;\tau)P_{pq}(q;\tau)G_{rc}(k|\tau,\tau'),
\label{eq:CLA_eq3}
\end{eqnarray}
where the operators $\widehat{\Sigma}_{abcd}$ and $\widehat{\Lambda}_{ab}$
are defined as
\begin{equation}
\widehat{\Sigma}_{abcd}(k;\tau)=\delta_{ac}\delta_{bd}\frac{\partial}{\partial \tau}+
\delta_{ac}\Omega_{bd}(k;\tau)+\delta_{bd}\Omega_{ac}(k;\tau),
\quad
\widehat{\Lambda}_{ab}(k;\tau)=\delta_{ab}\frac{\partial}{\partial \tau}+
\Omega_{ab}(k;\tau),
\end{equation}
The explicit expressions for the kernels $F_{apq}$ and $K_{cpq}$ are
summarized as
\begin{eqnarray}
&&F_{apq}(\bfk,\bfp,\bfq;\tau)=2\int_{\tau_0}^{\tau} d\tau''\,\Bigl[\,
2\,\,G_{ql}(q|\tau,\tau'')\,\gamma_{lrs}(\bfk,\bfp;\tau'')\,
R_{ar}(k;\tau,\tau'')R_{ps}(p;\tau,\tau'')
\nonumber\\
&&\quad\quad\quad\quad\quad\quad\quad\quad\quad\quad+\,\,\,
G_{al}(k|\tau,\tau'')\,\gamma_{lrs}(\bfp,\bfq;\tau'')\,
R_{pr}(p;\tau,\tau'')\,R_{qs}(q;\tau,\tau'')
\,\Bigr],
\label{eq:kernel_F2}\\
 &&K_{cpq}(\bfk',\bfp,\bfq;\tau,\tau')
 \nonumber\\
 &&\quad\quad\quad
 =\,\,4\int_{\tau_0}^{\tau} d\tau''\,
 \,G_{ql}(q|\tau,\tau'')\,\gamma_{lrs}(\bfk',\bfp;\tau'')\,
 R_{ps}(p;\tau,\tau'')
 \nonumber\\
 &&\quad\quad\quad\quad\quad\quad\quad\quad\times\,\,
 \Bigl\{ R_{cr}(k';\tau',\tau'')\Theta(\tau'-\tau'')+
 R_{rc}(k';\tau'',\tau')\Theta(\tau''-\tau')\Bigr\}
 \nonumber\\
 &&\quad\quad\quad\,+ \,\,2\int_{\tau_0}^{\tau'} d\tau''\,
 G_{cl}(k'|\tau',\tau'')\,\gamma_{lrs}(\bfp,\bfq;\tau'')\,
R_{pr}(p;\tau,\tau'')\, R_{qs}(q;\tau,\tau'').
\label{eq:kernel_K2}
\end{eqnarray}

The closure equations (\ref{eq:CLA_eq1})--(\ref{eq:CLA_eq3}) are
the integro-differential equations involving
several non-linear terms, in which the information of
the higher-order corrections in SPT is encoded. Thus, replacing all
statistical quantities in these non-linear terms with linear-order ones,
the solutions of the closure equations automatically reproduce the
leading-order results of SPT, so called one-loop power spectra.
Further, a fully non-linear treatment of the closure
equations provides a non-perturbative description of the power spectra,
and has an ability to predict the matter power spectra accurately beyond
one-loop SPT. Strictly speaking, the non-linear terms
in the right-hand side of Eqs.~(\ref{eq:CLA_eq1})--(\ref{eq:CLA_eq3}) have
only an information of the one-loop corrections.
However, it has been shown in Ref.~\cite{Taruya:2007xy} that the
present formulation are equivalent to the one-loop level of
renormalized perturbation theory by Crocce and Scoccimarro
\cite{Crocce:2005xy} and 2PI effective action method by Valageas
\cite{Valageas:2006bi}, and even the
leading-order approximation still contain some non-perturbative effects.
The application of the closure approximation, together with the detailed
comparison with N-body simulations, is presented in
Refs.~\cite{Nishimichi:2008ry,T2009}. Also,
comprehensive discussion on the differences between
several (semi-)analytic prescriptions for computing non-linear power
spectrum, including the closure approximation, is
found in Ref.~\cite{Carlson:2009it}.

In this paper, we mainly use the closure equations for the purpose of computing
the one-loop power spectra. The results for the fully non-linear treatment of
the closure equations will be presented elsewhere.
The numerical scheme to solve the closure equations is basically
the same as the one described in Ref.~\cite{HT2009}. In Appendix A, we
briefly review the numerical scheme and summarize several modifications.

Before closing this section, we note here that the resultant equations
(\ref{eq:CLA_eq1})--(\ref{eq:CLA_eq3}) contain the additional
non-linear terms originating from the modification of the Poisson equation
(see Eq.(\ref{eq:modified_Poisson})). In particular, the terms
containing the higher-order vertex function $\sigma_{abcd}$
can be effectively absorbed into the matrix $\Omega_{ab}$.
Since the non-vanishing contribution of the higher-order vertex function only
comes from $\sigma_{2111}$,
this means that the effective Newton constant $G_{\rm eff}$
defined by (\ref{eq:effective_G}) is renormalised as
$G_{\rm eff}\to G_{\rm eff}+\delta G_{\rm eff}$, with $\delta G_{\rm eff}$
given by
\begin{equation}
\delta G_{\rm eff} = \frac{3H^2}{4\pi\,\rho_{m}}
\int \frac{d^3\bfq}{(2\pi)^3}\,\sigma_{2111}(\bfq,-\bfq,\bfk;\tau)\,
P_{11}(q;\tau).
\end{equation}
This is a clear manifestation of the mechanism
to recover GR on small scales through the renormalisation of the Newton
constant and, due to this mechanism, the growth rate of
structure formation is altered on non-linear scales.

\section{Applications}

In this section, based on the formulations presented in
Sec.~\ref{sec:quasi-static_PT} and \ref{sec:closure_approx}, we
compute the one-loop power spectrum in specific models of
modified gravity theory. The models considered here are
DGP braneworld models and $f(R)$ gravity models, which we will in turn
discuss in Sec.~\ref{subsec:DGP} and \ref{subsec:fR}, respectively.

\subsection{DGP models}
\label{subsec:DGP}

In this subsection, we consider DGP braneworld models \cite{dvali00} as
a representative example of modified gravity models in
the context of higher-dimensional cosmology.
In this model, it is
possible to derive the quasi non-linear power spectrum analytically
as is discussed in details in Appendix B. This provides us
a check of our numerical code explained in the previous section
and Appendix A.

\subsubsection{DGP models}
In DGP models, we are supposed to be living in a 4D brane in a 5D
spacetime. The model is described by the action given by
\begin{equation}
S= \frac{1}{4 \kappa^2 r_c} \int d^4 x \sqrt{-g_5} R_5
+ \frac{1}{2 \kappa^2} \int d^4 x \sqrt{-g} (R  +L_{\rm m}),
\end{equation}
where $\kappa^2 = 8 \pi G$, $R_5$ is the Ricci scalar in 5D and $L_{\rm m}$
stands for the matter Lagrangian confined on a brane. The cross over scale
$r_c$ is the parameter in this model which is a ratio between the 5D Newton
constant and the 4D Newton constant. The modified Friedman equation is given by
\begin{equation}
\epsilon \frac{H}{r_c} = H^2 - \frac{\kappa^2}{3} \rho,
\end{equation}
where $\epsilon = \pm 1$ represents two distinct branches of the solutions
\cite{deffayet00}. From this modified Friedman equation, we find that the cross-over
scale $r_c$ must be fine-tuned to be the present-day horizon scales
in order to modify gravity only at late times.
The solution with $\epsilon = +1$ is known as the self-accelerating branch
because even without the cosmological constant, the expansion of the Universe
is accelerating as the Hubble parameter is constant, $H= 1/r_c$. On the
other hand $\epsilon = -1$
corresponds to the normal branch. In this branch, we need a cosmological
constant to realize the cosmic acceleration. However, due to the modified
gravity effects, the Universe
behaves as if it were filled with the Phantom dark energy with the equation
of state $w$
smaller than $-1$ \cite{Sahni:2002dx,Lue:2004za}.
It is known that the self-accelerating solution is plagued
by the ghost instabilities (see \cite{Koyama:2007za} for a review).
Also it gives a poor fit to the observations
such as supernovae and cosmic microwave background anisotropies
\cite{Fairbairn:2005ue, Maartens:2006yt, Fang:2008kc}.

However, this model is the simplest modified gravity model where the mechanism
to recover of GR on small scales is naturally encoded and it can
be used to get insights into the effect of this mechanism on the non-linear
power spectrum. Also in this model, it is possible to derive analytic
expressions for the quasi non-linear power spectrum. Thus it offers a check
of our numerical code to solve the evolution equations for the
power spectrum.

In this model, gravity becomes 5D on large scales larger than $r_c$.
On small scales, gravity becomes 4D but it is not described by GR.
The quasi-static perturbations are described by the BD theory where
the BD parameter is given by \cite{Koyama:2005kd}
\begin{equation}
\omega_{\rm BD}(\tau) = \frac{3}{2} \Big(\beta(\tau)-1 \Big),
\quad \beta(\tau)= 1 - 2 \epsilon H r_c
\left(1+ \frac{\dot{H}}{3 H^2} \right),
\end{equation}
where $\dot{H}$ is the cosmic time derivative of the Hubble parameter $H$.
Note that the BD parameter depends on time in this model.
The BD scalar is massless $M_1=0$. However, it acquires
a large second order interaction given by \cite{Koyama:2007ih}
\begin{equation}
{\cal I}(\varphi) = \frac{r_c^2}{a^4}
\Big[(\nabla^2 \varphi)^2 -(\nabla_i \nabla_j \varphi)^2 \Big].
\end{equation}
Note that $r_c$ is tuned to be the present-day horizon scale. Thus
this second order term has a large effect. The higher order terms
than the second order are suppressed by the additional powers of the
4D Planck scale and, in the Newtonian limit,
we can safely ignore them. Therefore, in this model, we have
\begin{eqnarray}
M_1=0, \quad M_2(\bfk_1,\bfk_2) &=& 2 \frac{r_c^2}{a^4} \Big[
k_1^2 k_2^2 - (\bfk_1 \cdot \bfk_2)^2
\Big], \quad M_3 =0, \\
\Pi(k,\tau) &=& \beta(\tau) \frac{k^2}{a^2}.
\end{eqnarray}
In this paper, we use the best-fit cosmological
parameters for the flat self-accelerating
universe: $\Omega_{m}=0.258, \Omega_{b}= 0.0544, h=0.66, n_s=0.998$
\cite{Fang:2008kc}.
For the normal branch, we add the cosmological constant $\Omega_{\Lambda}=1.5$.

It has been suggested that the idea
of the self-acceleration could be extended to more general models and
in these models one could avoid the ghost instabilities \cite{Nicolis:2008in}.
Although the covariant theory that realizes this idea is not known yet but
our formalism can be applied to calculate the non-linear power
spectrum in these models. In fact, in our formalism, this new
model corresponds to add a new term $M_3$ (and also
a constant term in ${\cal I}$). We will leave it as a future work
to study these extensions.

\subsubsection{Quasi non-linear power spectra}

\begin{figure}[h]
\centerline{
\includegraphics[width=17cm]{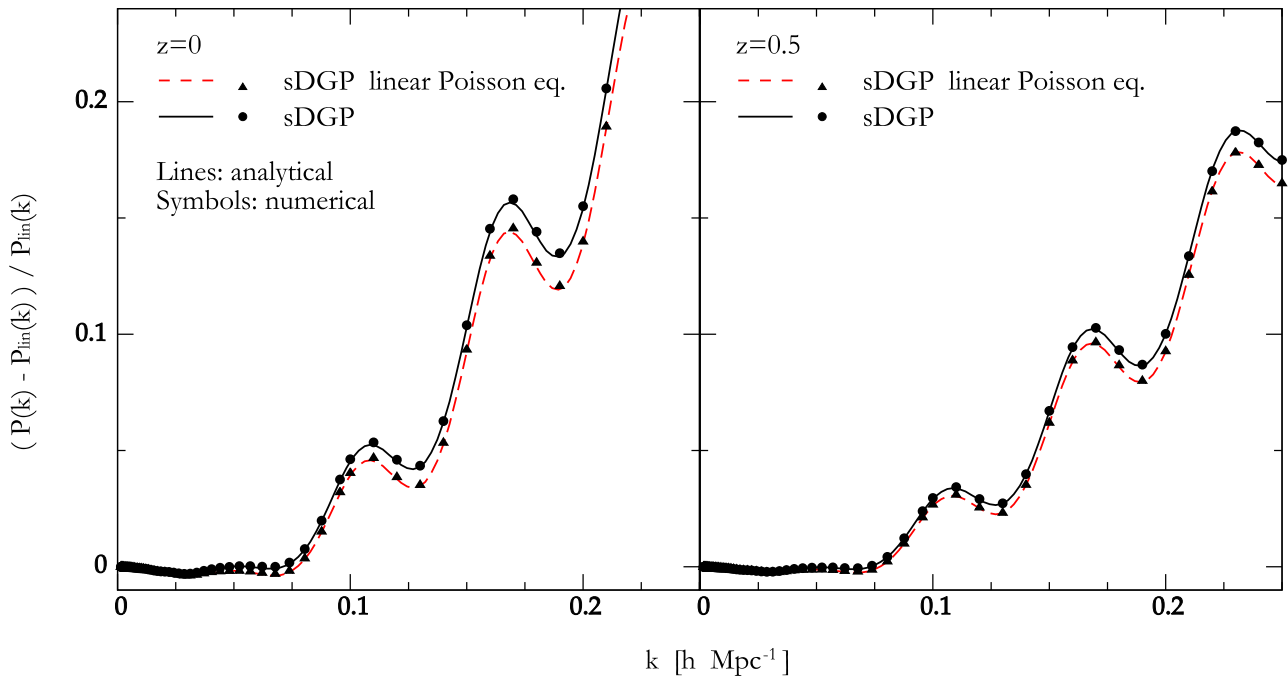}}
\caption{Fractional change in the non-linear power spectrum
relative to the linear power spectrum in the self-accelerating branch of
DGP. The solid (black) line  is the analytic solutions and the circles are
numerical solutions obtained by solving the closure equation.
The dashed (red) line shows the analytic solutions obtained by neglecting
the non-linear interaction terms ${\cal I}$. The triangles represent
the numerical solutions in this case. We used the best fit cosmological
parameters for the flat universe $\Omega_{m}=0.258, \Omega_{b}= 0.0544, h=0.66, n_s=0.998$.}
\label{fig1}
\end{figure}
\begin{figure}[h]
\centerline{
\includegraphics[width=17cm]{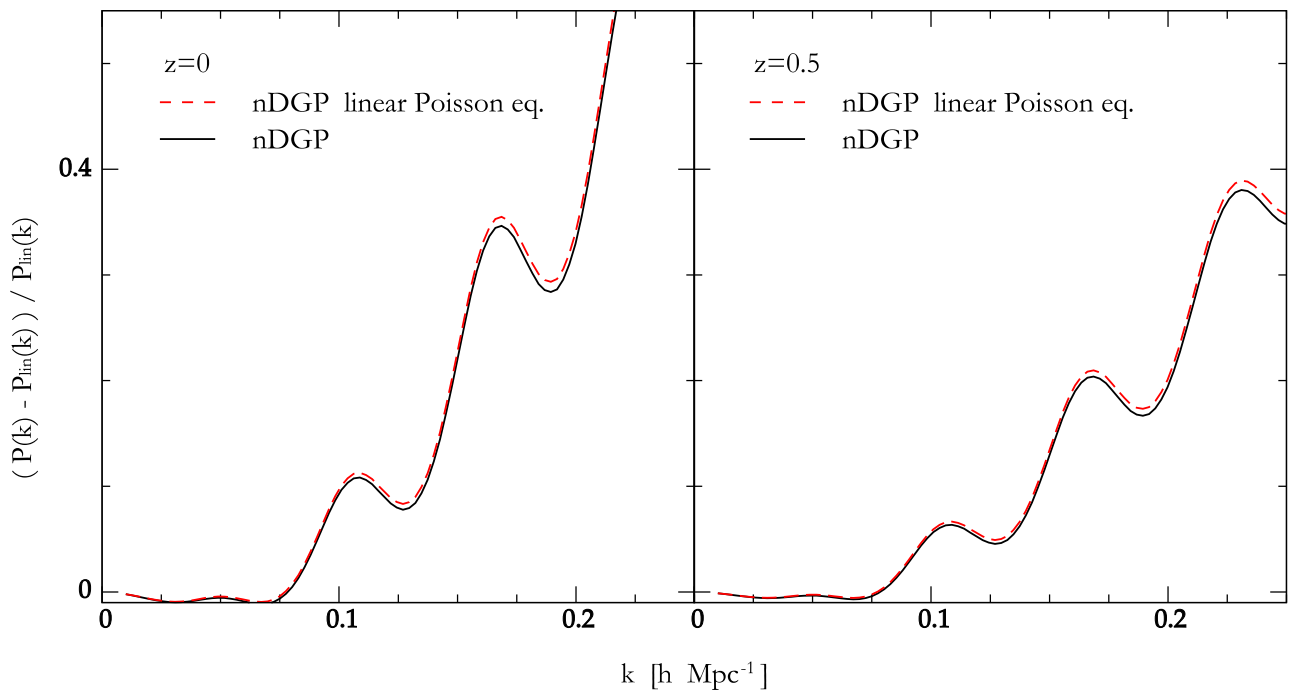}}
\caption{The same in the normal branch as in Fig.1.
We only show the analytic solutions. The numerical
solutions agree with them very well. The cosmological
parameters are the same as the self-accelerating universe but
in addition there is a cosmological constant $\Omega_{\Lambda}=1.5$.}
\label{fig2}
\end{figure}

Fig.~1 shows a fractional change in the non-linear power spectrum for density perturbations
relative to the linear power spectrum in the self-accelerating branch.
In order to see the effect of the non-linear interaction term
${\cal I}$, we also plotted the non-linear power spectrum obtained from
the linear Poisson equation by neglecting the interaction term
${\cal I}$. Although the differences between the two spectra
are small in the quasi non-linear regime,
we can see that the non-linear interaction term ${\cal I}$
enhances the non-linear power spectrum. This is natural because
in the self-accelerating branch, the linear growth rate is suppressed compared
to the GR model that follows the same expansion history. The suppression is due to
the negative BD parameter $\omega_{\rm BD}<0$ in this model, which makes the Newton constant
smaller than GR. This is closely related to the fact that the BD scalar
becomes a ghost. Classically, the ghost mediates
a repulsive force and suppresses the gravitational collapse.
The non-linear interaction makes the theory approach GR. Thus it effectively
increases the Newton constant by screening the BD scalar. Then the power spectrum
receives an enhancement compared with the case without the non-linear interaction.

Fig.~2 shows a fractional change in the non-linear power spectrum
relative to the linear power spectrum in the normal branch. Again
we showed the two cases with and without ${\cal I}$. In the normal
branch, the linear growth rate is enhanced as the BD parameter
is positive $\omega_{\rm BD}>0$. Thus, the situation is completely
opposite to the self-accelerating branch and the non-linear
interaction suppresses the non-linear power spectrum in order to
recover GR on small scales.

In both branches, it is possible to derive the solutions
for the power spectrum analytically by solving the
equation for perturbations (\ref{eq:basic_eq}) perturbatively
\begin{equation}
\Phi_a = \Phi^{(1)}_a + \Phi^{(2)}_a + \Phi^{(3)}_a + ... \end{equation}
The power spectrum is also expanded accordingly
\begin{equation}
P_{ab}(k;t) = P_{ab}^{(11)}(k;t) + P_{ab}^{(22)}(k;t) +
P_{ab}^{(13)}(k;t) + ... \end{equation}
The detailed calculations are summarized in Appendix B.
In Fig.1, we compare the results obtained by solving the
closure equations numerically with those from the analytic solutions.
In order to derive the analytic solutions, we employed the
Einstein-de Sitter (EdS) approximation. In the EdS approximation,
all the non-linear growth rates appearing in the higher-order
solutions are approximately determined by the linear growth rate $D_1(t)$.
It is also possible to apply the EdS approximation in the numerical calculations
\cite{HT2009} and we have checked that the EdS approximation changes the result only
at sub percent level. The fact that the two results agree very well confirms
the validity of our numerical code.

\subsection{$f(R)$ gravity models}
\label{subsec:fR}

In this subsection, we derive the quasi non-linear power spectrum
in $f(R)$ gravity model (see \cite{Sotiriou:2008rp, Capozziello:2007ec}
for reviews).
In this model, N-body simulations have
been performed \cite{Oyaizu:2008sr, Oyaizu:2008tb, Schmidt:2008tn}
and we will check our numerical solutions against the full N-body
simulations.

\subsubsection{$f(R)$ gravity models}

We consider another class of modified theory of gravity that generalizes the
Einstein-Hilbert action to include an arbitrary function of
the scalar curvature $R$:
\begin{equation}
S=\int d^4x \sqrt{-g}\left[\frac{R+f(R)}{2 \kappa^2} +L_{\rm m}\right],
\end{equation}
where $\kappa^2=8\pi\,G$ and $L_{\rm m}$ is the Lagrangian of the ordinary
matter. This theory is equivalent to the BD theory with $\omega_{\rm BD}=0$ but
there is a non-trivial potential
\cite{Nojiri:2003ft, Chiba:2003ir}. This can be seen from the trace of modified
Einstein equations:
\begin{equation}
3 \Box f_R - R + f_R R -2 f = - \kappa^2 \rho,
\end{equation}
where $f_R = df/dR$ and $\Box$ is a Laplacian operator and we assumed
matter dominated universe.
We can identify $f_R$ as the BD scalar field and its
perturbations are defined as
\begin{equation}
\varphi = \delta f_R\equiv f_R-\overline{f}_R,
\end{equation}
where the bar indicates that the quantity is evaluated on the
cosmological background. In this paper, we assume $| \bar{f}_R | \ll 1$
and $|\bar{f}/\bar{R}| \ll 1$. These conditions are necessary to have the background
close to $\Lambda$CDM cosmology. Then the BD scalar perturbations satisfy
\begin{equation}
3 \frac{1}{a^2} \nabla^2 \varphi = - \kappa^2 \rho_m \delta
+ \delta R,
\quad \delta R \equiv R(f_R)-R(\overline{f}_R).
\end{equation}
This is nothing but the equation for the BD scalar perturbations
with $\omega_{\rm BD}=0$ and the potential gives the non-linear interaction term
\begin{equation}
{\cal I}(\varphi)= \delta R(\varphi).
\end{equation}
Then we find
\begin{eqnarray}
&&M_1= \overline{R}_f(\tau) \equiv \frac{d \overline{R}(f_R)}{d f_R},\quad
M_2 = \overline{R}_{ff}(\tau) \equiv \frac{d^2\overline{R}(f_R)}{d f_R^2},\quad
M_3 = \overline{R}_{fff}(\tau) \equiv \frac{d^3\overline{R}(f_R)}{d f_R^3},
\nonumber\\
&&\Pi(k,\tau)=\left(\frac{k}{a}\right)^2+\frac{\overline{R}_f(\tau)}{3}.
\nonumber
\end{eqnarray}
We should note that in this model, the linear growth rate depends on
the wave number. Due to this, the vertex functions are not the
separable functions in terms of $k$ and $\tau$. This prevents us
deriving the solutions analytically unlike the DGP case and we
need to solve the closure equation directly.

In this paper, we consider the function $f(R)$ of the form
\begin{equation}
f(R) \propto \frac{R}{A R +1},
\label{model}
\end{equation}
where $A$ is a constant with dimensions of length squared
\cite{Hu:2007nk}.
In the limit $R \to 0$, $f(R) \to 0$ and there is no
cosmological constant. For high curvature $A R \gg 1$,
$f(R)$ can be expanded as
\begin{equation}
f(R) \simeq - 2 \kappa^2 \rho_{\Lambda} - f_{R0} \frac{\bar{R}_0^2}{R},
\end{equation}
where $\rho_{\Lambda}$ is determined by $A$, $\bar{R}_0$
is the background curvature today and we defined $f_{R0}$ as
$f_{R0}=\bar{f}_R(R_0)$. As we mentioned before, we take
$|f_{R0}| \ll 1$ and assume that the background expansion
follows the $\Lambda$ CDM history with the same $\rho_{\Lambda}$.
The $M_1$ term determines the mass of the BD field  $m_{\rm BD}
= (M_1/3)^{1/2}$ as
\begin{equation}
m_{\rm BD}(\tau) \equiv \sqrt{\frac{\bar{R}_f}{3}} = \left(\frac{R_0}{6 |\bar{f}_R|}
\sqrt{\frac{f_{R0}}{\bar{f}_R}}\right)^{1/2} .
\end{equation}
Above the Compton length $m_{\rm BD}^{-1}$, the BD scalar
interaction decays exponentially and we recover GR. On small scales, we
recover the BD theory with $\omega_{\rm BD}=0$ if we neglect the
higher order terms $M_i, (i>1)$.
From Eq.~(\ref{eq:effectiveG}),
the Newton constant is $4/3$ times large
than GR. Thus the linear power spectrum acquires a scale dependent
enhancement on small scales. Of course, this model is excluded from
local gravity constraints. The higher order terms $M_i (i>1)$
are responsible for suppressing this modification of gravity on
small scales via the chameleon mechanism and it makes possible to
pass local gravity constraints. Thus we expect that
the non-linear interaction terms ${\cal I}$ will suppress the non-linear
power spectrum. In the following, when we mention the power spectrum
with the chameleon mechanism, it means that we introduce the non-linear
terms ${\cal I}$ derived from (\ref{model}) which contains the mechanism
to recover GR on small scales by screening the scalar mode.
In this paper, we adopt the cosmological parameters
given by $|f_{R0}|=10^{-4}, n_s=0.958, \Omega_{m}=0.24,
\Omega_{b}=0.046, \Omega_{\Lambda}=0.76, h=0.73$.

In this model, the solar system constraints are satisfied but it
has been pointed out that the chameleon mechanism does not work for
strong gravity and neutron stars cannot exist
\cite{Frolov:2008uf,Kobayashi:2008tq}. It was claimed that
a fine-tuned higher
curvature corrections to $f(R)$ is needed to cure the problem
\cite{Kobayashi:2008wc}. However, recently there appeared
an objection against the absence of relativistic stars \cite{Babichev:2009td}.
In this paper, we perturbatively take into account the chameleon
mechanism in the cosmological background and the quasi non-linear
power spectrum would be insensitive to the higher curvature
corrections.

We should also mention that recently a couple of papers appear that
study the second order perturbations and three point functions
in $f(R)$ gravity models \cite{Tatekawa:2008bw, Borisov:2008xn}.

\subsubsection{Quasi non-linear power spectra}

\begin{figure}[h]
\centerline{
\includegraphics[width=17cm]{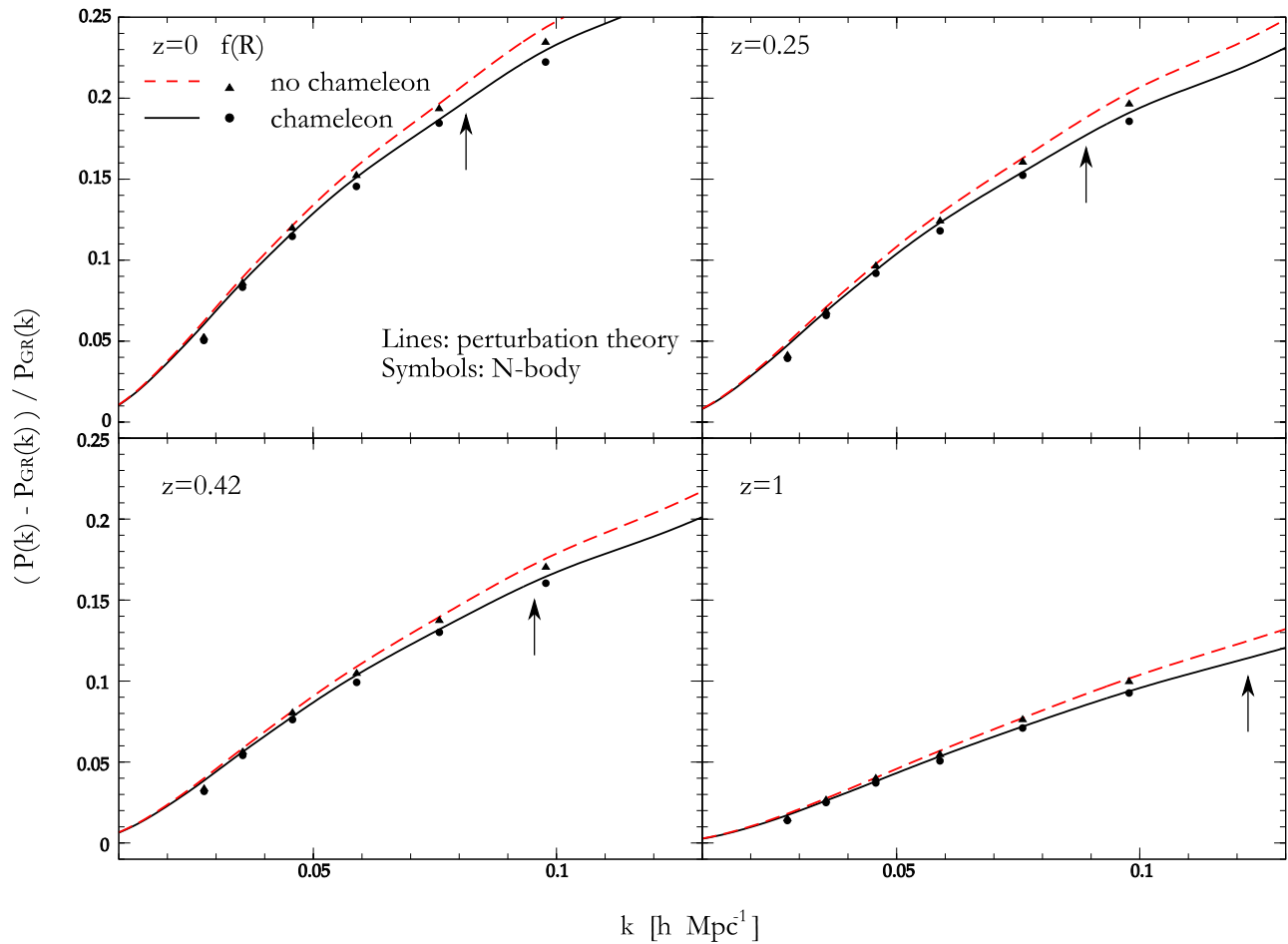}} \caption{Fractional change in the non-linear power spectrum
in $f(R)$ gravity models relative to the GR models with the same expansion
history. $P_{\rm GR}$ is the non-linear power
spectrum in $\Lambda$CDM model with the same cosmological parameters.
The solid (black) lines are the solutions in the perturbation theory obtained
by solving
the closure equation numerically. The circles show the results of N-body
simulations. The dashed (red) line is the perturbation theory solutions
obtained by neglecting the non-linear interaction terms ${\cal I}$.
The triangles represent the corresponding N-body solutions. The arrow
indicate the valid regime of the perturbation theory.
The parameters are taken as $|f_{R0}|=10^{-4}, n_s=0.958, \Omega_{m}=0.24,
\Omega_{b}=0.046, \Omega_{\Lambda}=0.76, h=0.73$.
}
\label{fig3}
\end{figure}

Fig.~3 shows a fractional change in the non-linear power spectrum
in $f(R)$ gravity models relative to the GR model with the same
expansion history, $\Lambda$CDM model.
Note again that unlike the DGP case, the function $\Pi$ is not a separable
function of $k$ and $\tau$, and the analytical calculation
is intractable.
The solid line shows the case with the chameleon mechanism
which includes the higher order terms $M_2$ and $M_3$. The dashed line
shows the non-chameleon case where we neglected the higher order
terms $M_2$ and $M_3$. As expected, the chameleon mechanism
suppresses the non-linear power spectrum. The situation is similar
to the normal branch DGP models.

We have also shown the results from N-body simulations in
both cases \cite{N-body}. The N-body simulations are performed under
the quasi-static approximations. The form of $f(R)$ that is used in
the simulations is the same as ours and we have checked that the
basic equations that are used in their simulations are the same
as our equations. They run the simulations
with a box size $256 h^{-1}$ Mpc and with $256^3$ particles.The
power spectrum starts to show systematic $>10 \%$ deviations
from Smith et.al. fitting formula \cite{Smith:2002dz}
for $k >0.79 h$ Mpc $^{-1}$.
The arrow in the figure shows the expected validity
range of the perturbation theory. We determine this validity
regime of the perturbation theory by solving the equation
\begin{equation}
\frac{k^2}{6 \pi^2} \int^{k}_0 d^3 q P_{\rm lin}(q,z) =0.18,
\label{klim}
\end{equation}
for $k$, where $P_{\rm lin}$ is the linear power spectrum.
This limit of $k$ represents the range of the 1$\%$-level accuracy,
which has been empirically found by comparisons between
the perturbation theory predictions and N-body simulations \cite{Nishimichi:2008ry}.
Of course, this condition was calibrated in simulations in GR and
there is no guarantee that this condition can be applied to modified gravity
models, but this limit gives an useful indication of the
validity of the perturbations theory. In fact, we find that
the agreement between the perturbations theory and N-body
simulations in $f(R)$ theory is good within this validity
regime of the perturbation theory.

\section{Implication for fully non-linear power spectrum}
We have developed the formalism to derive the quasi non-linear
power spectrum. In this section, we study the
Parametrized Post-Friedmann (PPF) framework for the non-linear
power spectrum \cite{Hu07} in order to get insight into the fully non-linear
power spectrum.

\subsection{PPF formalism}
Hu and Sawicki proposed a fitting formula for the non-linear power
spectrum in modified gravity models \cite{Hu07}.
The fitting formula based on the
observation that the non-linear power spectrum should approach the
one in the GR model that follows the same expansion history of the
Universe due to the recovery of GR on small scales. They postulate
that the fully non-linear power spectrum in a modified gravity model
is given by the formula
\begin{equation}
P(k,z)=\frac{P_{\rm non-GR}(k,z)+c_{\rm nl} \Sigma^2(k,z)
P_{\rm GR}(k,z)}{1+c_{\rm nl} \Sigma^2(k,z)},
\label{PPF}
\end{equation}
where $z$ is a red-shift.
Here $P_{\rm non-GR}(k,z)$ is the non-linear power spectrum which
is obtained without the non-linear interactions that are
responsible for the recovery of GR. This is equivalent to
assume that gravity is modified down to the small scales
in the same way as in the linear regime. $P_{\rm GR}(k,z)$ is the non-linear
power spectrum obtained in the GR dark energy model that follows
the same expansion history of the Universe as the modified
gravity model. The function $\Sigma^2(k,z)$ determines
the degree of non-linearity at a relevant wavenumber $k$.
They propose to take $\Sigma^2(k) = k^3 P_{\rm lin}(k,z)/2 \pi^2$,
where $P_{\rm lin}(k,z)$ is the linear power spectrum in the
modified gravity model. Finally, $c_{\rm nl}$ is a parameter in this
framework which controls the scale at which the theory approaches GR.

Once we obtain the quasi non-linear power spectrum, we can
check whether the PPF framework works and determines $c_{\rm nl}$
in the quasi non-linear regime. In our formalism, $P_{\rm non-GR}(k,z)$ is
obtained by neglecting the non-linear interaction ${\cal I}$. $P_{\rm GR}(k,z)$
can be obtained by taking $\omega_{\rm BD} \to \infty$ limit and also neglecting
${\cal I}$. We again consider two explicit examples.

\subsection{DGP models}
We first consider the self-accelerating branch solutions in DGP
models. The extension to the normal branch solutions
is straightforward though there is a subtlety in defining
the equivalent dark energy model as the equation
of state of dark energy becomes less than $-1$ and
it diverges at some redshift \cite{Lazkoz:2006gp}. Also
an addition of curvature in the background is necessary to have
large modified gravity effects \cite{Giannantonio:2008qr}.
We leave this for a future work.
In the self-accelerating universe, we find that
\begin{equation}
\Sigma^2(k,z) = \frac{k^3}{2 \pi^2} P_{\rm lin}(k,z),
\end{equation}
as proposed by Hu and Sawicki \cite{Hu07} gives a nice fit to the results
obtained in the perturbation theory. We find that by allowing the time
dependence
in $c_{\rm nl}$, it is possible to recover the solutions for the non-linear
power spectrum very well within the validity regime of the perturbation
theory determined by Eq.~(\ref{klim}). At $z=0$, $c_{\rm nl}$ is given by
0.3 and there is a slight redshift dependence (Fig.~4). For
$0 \leq z \leq 1$, we find that $c_{\rm nl}$ can be approximately
fitted as $c_{\rm nl} = 0.3(1+z)^{0.16}$.

Armed with this result, it is tempting to extend our analysis to the fully
non-linear regime. In GR, there are several fitting formulae which provide
the mapping between the linear power spectrum and non-linear power spectrum.
It is impossible to apply these mapping formulae to modified gravity
models as the mapping does not take into account the non-linear interaction
terms ${\cal I}$ in the Poisson equation.
If we apply the GR mapping formula to the linear power spectrum, we
would get the non-linear power spectrum without ${\cal I}$, that is,
$P_{\rm non-GR}(k)$.
In fact, there exists N-body simulations in DGP models performed by
neglecting the non-linear interaction terms.

The mapping formulae should be valid in GR models, so we can also predict
$P_{\rm GR}(k)$. Then using the PPF formalism (\ref{PPF}), we can predict
the non-linear power spectrum if $c_{\rm nl}$ is known. In the right panel of
Fig.~4, we plotted a fractional change in the power spectrum
in the DGP model relative to the GR model with the same expansion history.
We used the fitting formula developed by Smith et.al. \cite{Smith:2002dz}.
It should be noted that this fitting formula should be used with caution
even in the simple case of dynamical dark energy \cite{Casarini:2008np}
but given a lack of better fitting formula available at the moment,
we used this formula as an example.
If we could extrapolate the result in the quasi non-linear regime, we
would have $c_{\rm nl}=0.3$ at $z=0$.

Recently, N-body simulations in DGP models have been done by Schmidt
\cite{Schmidt:2009sg}. Fig.5 shows the comparison between the PPF prediction
and N-body simulations. In the left panel of Fig.~5, the dashed line shows
the power spectrum with
the linear Poisson equation without ${\cal I}$. The corresponding N-body
results are shown by triangles. Although Ref.~\cite{Laszlo:2007td}
reported that the fitting formula works fine in this case,
we find that the fitting formula by Smith.et.al. slightly overestimates
the power spectrum \cite{Schmidt:2009sg}.
The solid curve shows the full power spectrum including ${\cal I}$. Again
the PPF formalism slightly overestimates the power spectrum. If we take the
ratio between the two cases, the PPF formalism recovers N-body results up to
$k =0.5$Mpc h$^{-1}$. The validity regime of perturbation theory is below
$k =0.12$Mpc h$^{-1}$. Thus the PPF formalism using $c_{\rm nl}$
derived by the perturbation theory describes the effect
of the Vainshtein mechanism on non-linear scales beyond the
validity regime of the perturbation theory.

This observation suggests that an improvement of the PPF
formalism can be made by getting a more accurate power spectrum
without the Vainshtein mechanism because the
PPF formalism describes the effect of the Vainshtein mechanism
very well. In order to
demonstrate this fact, we derive the power spectrum without
the Vainsgtein mechanism $P_{\rm non-GR}$ by interpolating the N-body results
(see the right panel of Fig.~5).
Using this power spectrum as the power
spectrum with the linear Poisson equation $P_{\rm non-GR}$ in the PPF formalism,
we find that the full power spectrum can be very well described by the
PPF formalism where $c_{\rm nl}$ is derived by the perturbation
theory. We should emphasize that the ratio between the
power spectra with and without the chameleon mechanism is not very sensitive
to the power spectrum with the linear Poisson equation. This also indicates that
the PPF formalism with $c_{\rm nl}$ determined by the
perturbation theory describes the effect of the Vainshtein
mechanism very well at least up to $k \sim 0.5h$Mpc$^{-1}$.

Above $k=1$Mpch$^{-1}$, N-body simulations do not have enough
resolutions.
If we can extrapolate our results to this regime, we find that even
at $k=10$Mpch$^{-1}$, the difference between the power spectrum in DGP
and that in the equivalent GR model remains at $7\%$ level. This is crucial
to distinguish between the two models using weak lensing as the
signal to noise ratio is larger on smaller scales. Of course,
we should emphasize that there is no guarantee that $c_{\rm nl}$
measured in the quasi non-linear regime is valid down to the
fully non-linear scales and this should be tested by N-body simulations
with higher resolutions.

\begin{figure}[h]
\centerline{
\includegraphics[width=18cm]{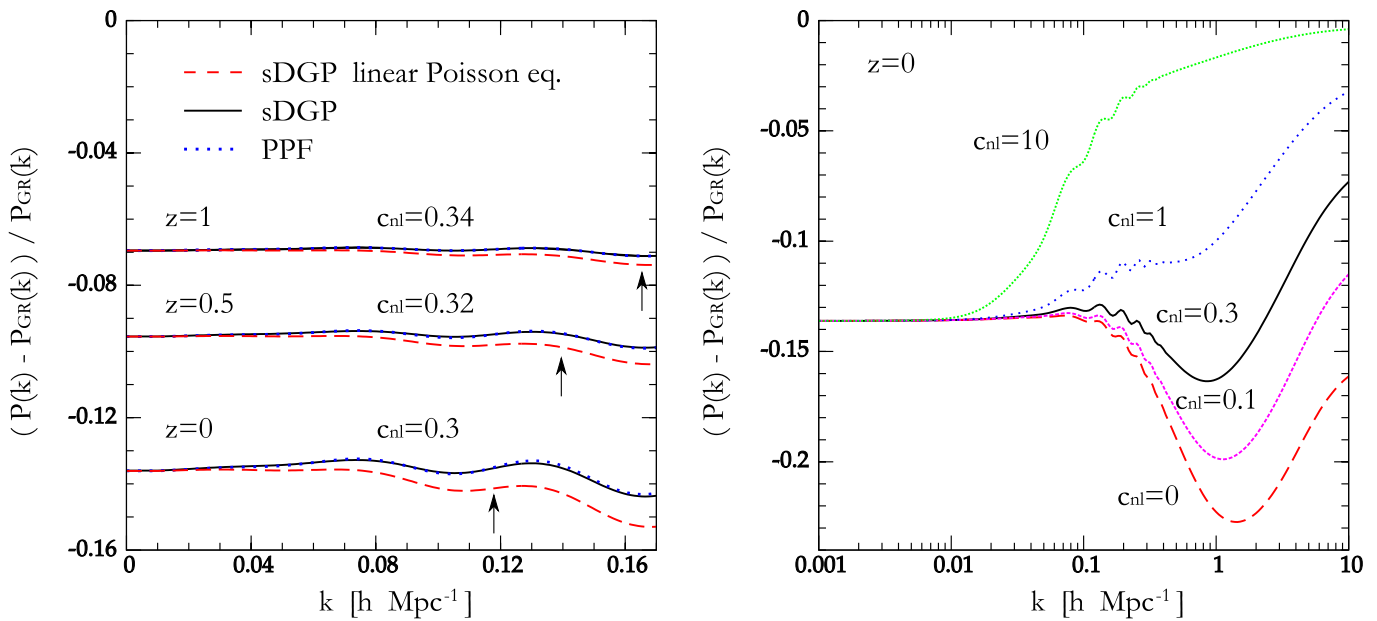}}
\caption{A fractional change in the power spectrum
in the DGP self-accelerating solution relative to the GR model which has the same
expansion history as the DGP. The solid (black) line shows the perturbation
theory solution and the dashed (red) line shows the perturbation theory
solution without the non-linear interaction terms in the Poisson equation.
The dotted (blue) line shows the PPF fitting. By allowing the redshift dependence
of $c_{\rm nl}$, we can fit the power spectrum very well within the
validity regime of the perturbation theory indicated by arrows.
The right panel shows the results at $z=0$ obtained from
the fitting formula by Smith et.al. for $P_{\rm non-GR}$ and $P_{\rm GR}$.
If $c_{\rm nl}=0.3$ obtained by the perturbation theory is applicable,
the solid (black) line is our prediction on non-linear scales.
The cosmological parameters are the same as in Fig~1.
}
\label{fig4}
\end{figure}

\begin{figure}[t]
\centerline{
\includegraphics[width=18cm]{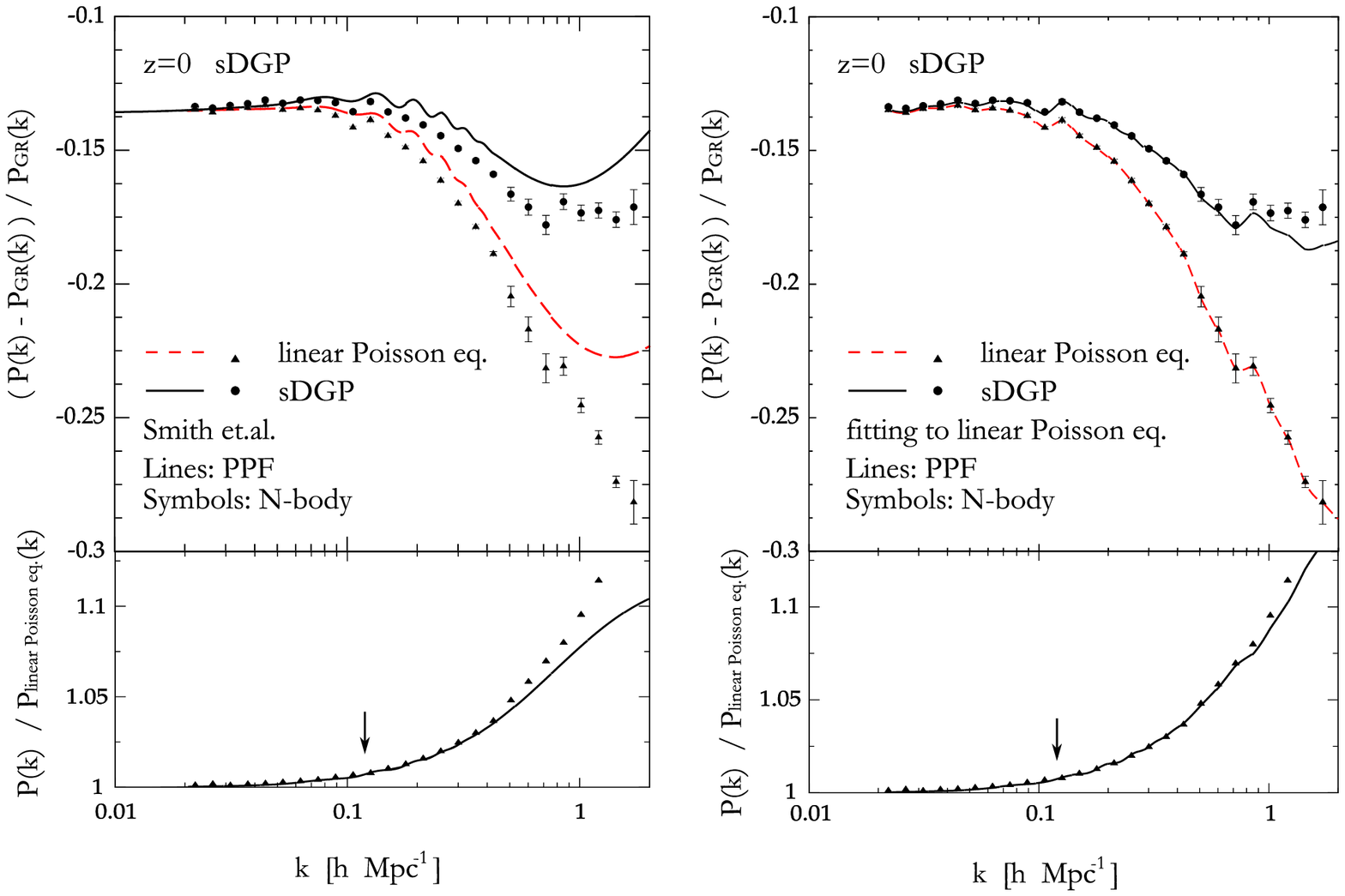}}
\caption{Comparison between the PPF prediction and N-body
simulations. In the left panel, Smith et.al. fitting formula
is used to predict $P_{\rm non-GR}$ and $P_{\rm GR}$. We used
$c_{\rm nl}$ determined by the perturbation theory
$c_{\rm nl}=0.3$ at $z=0$. In the right panel, we fitted
N-body results with the linear Poisson equation to derive
$P_{\rm non-GR}$.
}
\label{fig5}
\end{figure}
\subsection{$f(R)$ gravity models}
Next, we consider $f(R)$ gravity models.
As in the DGP model, we first check if
we can reproduce the perturbation theory results by the PPF fitting. We find
that the fitting is not very good if we adopt $\Sigma(k,z)$ as is proposed
by Hu and Sawicki \cite{Hu07}. Instead, if we choose $\Sigma(k,z)$ as
\begin{equation}
\Sigma^2(k,z) = \left( \frac{k^3}{2 \pi^2} P_{\rm lin}(k,z) \right)^{1/3},
\end{equation}
the solutions in the perturbation theory are fitted by the PPF formalism
very well by allowing the redshift dependence in $c_{\rm nl}$.
At $z=0$, $c_{\rm nl}=0.085$
gives an excellent fit to the power spectrum within the validity
regime of the perturbation theory. For $0 \leq z \leq 1$, we find that
$c_{\rm nl}$ can be approximately fitted as $c_{\rm nl} = 0.08(1+z)^{1.05}$.
In Fig.~6, we also show the prediction
for the fractional difference between the power spectrum in $f(R)$ theory
and that in $\Lambda$CDM model in fully non-linear regime for several
$c_{\rm nl}$.

It is also possible to check our predictions
against the N-body simulations done by
Refs.~\cite{Oyaizu:2008sr, Oyaizu:2008tb, Schmidt:2008tn}.
Fig.~7 shows the comparison between
the PPF prediction and N-body simulations. In the left panel,
the dashed line corresponds
to non-chameleon case with $c_{\rm nl}=0$. The corresponding N-body
results are shown by triangles. We again used the fitting
formula by Smith et.al. to derive the non-linear power spectrum from the
linear power spectrum. Compared with the N-body results, we find that,
in this case, the formula by Smith et.al.
slightly underestimates the power spectrum around $0.03h$Mpc$^{-1}< k
< 0.5h$Mpc$^{-1}$  and overestimates the power at $k > 0.5h$Mpc$^{-1}$
though N-body simulations have large errors in this regime.
The solid line shows the case with the chameleon mechanism. Again the
PPF formalism underestimates the power spectrum in the same region
as the non-chameleon case. If we take the ratio between the
non-chameleon case and chameleon case, the PPF formalism nicely
recovers the N-body results up to $k \sim 0.5h$Mpc$^{-1}$. Beyond
that, N-body simulations have large errors.
We should emphasize that the perturbation theory is valid only up to
$k =0.08h$Mpc$^{-1}$ at $z=0$. Thus the PPF formalism using $c_{\rm nl}$
derived by the perturbation theory describes the effect
of the chameleon mechanism on non-linear scales beyond the
validity regime of the perturbation theory.

As we have done in DGP models, we also derived the power spectrum without
the chameleon mechanism $P_{\rm non-GR}$ by interpolating the N-body results
(see the right panel of Fig.~7).
Using this power spectrum as the non-chameleon power
spectrum $P_{\rm non-GR}$ in the PPF formalism, we find that the power spectrum
with the chameleon mechanism can be very well described by the
PPF formalism where $c_{\rm nl}$ is derived by the perturbation
theory.
Again for larger $k$, N-body simulations also do not have enough resolutions
and it is difficult to tell whether this extrapolation
is good or not. More detailed study is needed to address
the power spectrum at larger $k$, but the PPF formalism
is likely to give a promising way to develop a fitting
formula for the non-linear power spectrum in modified
gravity models.

\begin{figure}[h]
\centerline{
\includegraphics[width=18cm]{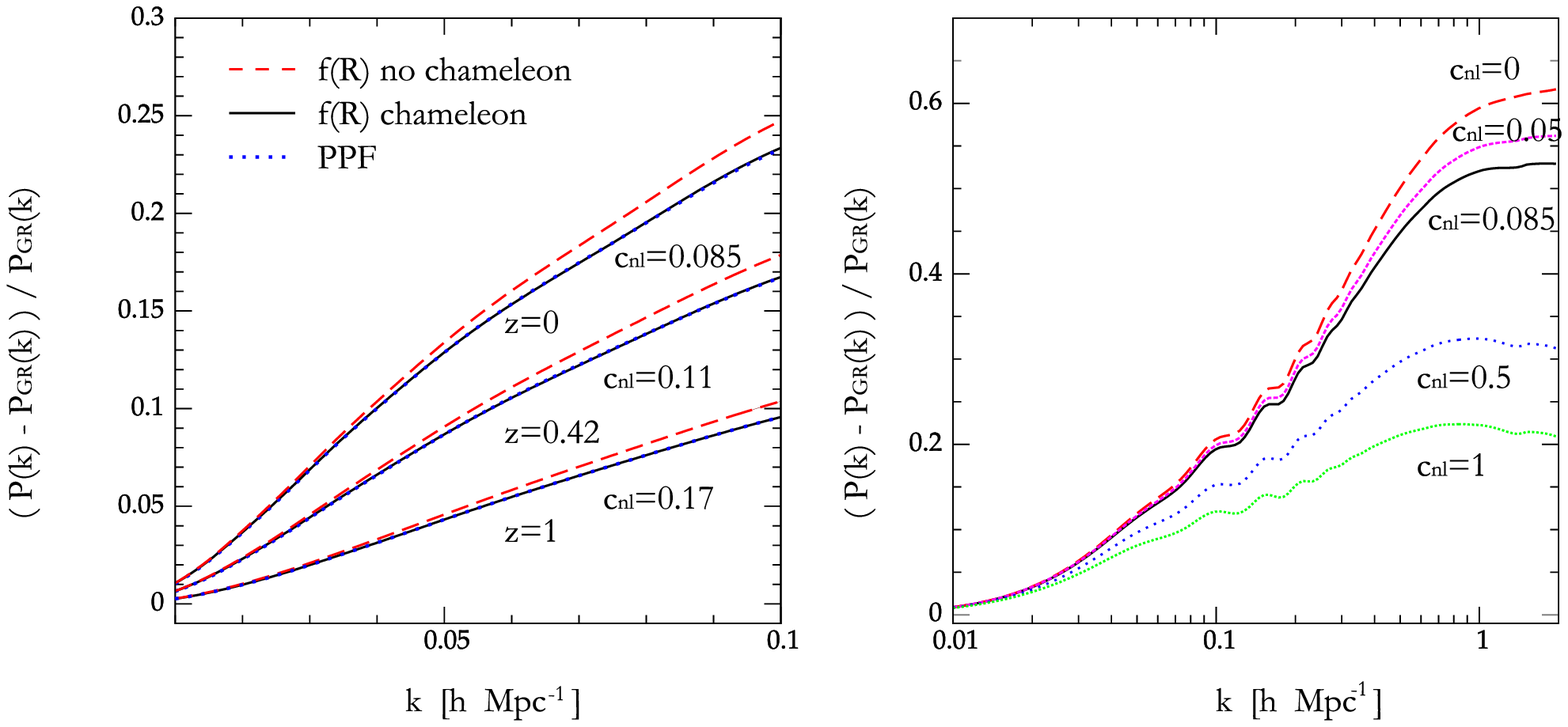}}
\caption{The same as Fig.~4 in $f(R)$ gravity models
}
\label{fig6}
\end{figure}
\begin{figure}[t]
\centerline{
\includegraphics[width=18cm]{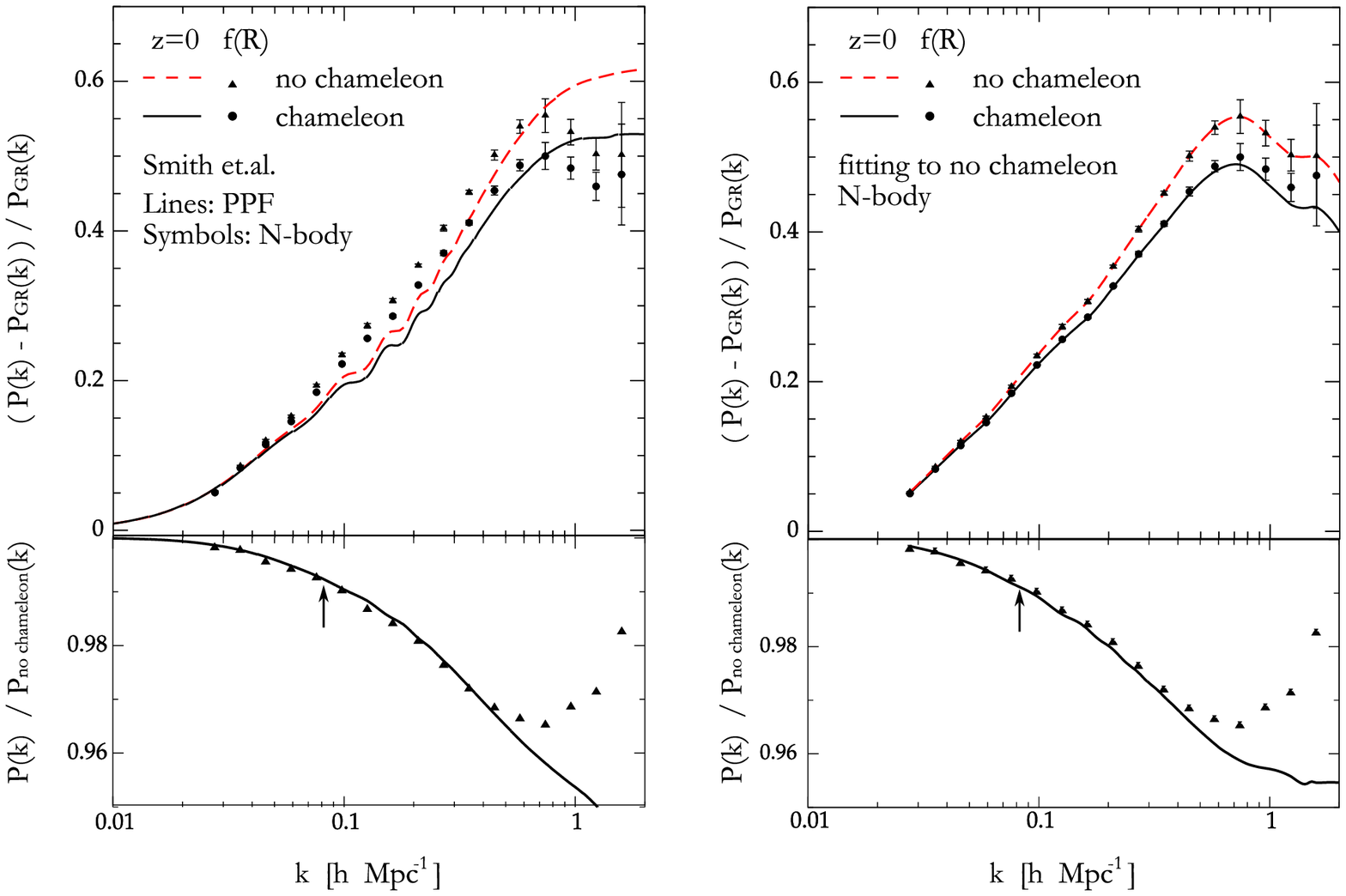}}
\caption{Comparison between the PPF prediction and N-body
simulations in $f(R)$ gravity models. In the left panel,
Smith et.al. fitting formula
is used to predict $P_{\rm non-GR}$ and $P_{\rm GR}$. We used
$c_{\rm nl}$ determined by the perturbation theory
$c_{\rm nl}=0.085$ at $z=0$. In the right panel, we fitted
N-body results without the chameleon mechanism to derive
$P_{\rm non-GR}$.
}
\label{fig6}
\end{figure}

\section{Conclusions}
In this paper, we developed a formalism to derive the quasi-nonlinear
power spectrum for density perturbations in modified gravity models.
We assume that the quasi-static perturbations under the horizon
scale can be described by a BD theory with a time dependent BD
parameter.
The non-linear interaction terms are model-dependent so we parametrised
these terms. Then we took into account the non-linear
interaction terms perturbatively. The Poisson equation
becomes a non-linear equation which relates the curvature perturbation
to the density perturbations non-linearly. Then combining it to
the continuity equation and Euler equation assuming the
irrotationality of fluid quantities, we derived the evolution
equations for the density perturbations and the velocity divergence
in a compact form. The non-linearity of the Poisson equation
introduces new vertex functions.

Then we derived the evolution equations for the power spectrum.
The closure approximation was employed to derive a
closed set of equations.
In this paper, we further simplified the equations by replacing all the quantities
in non-linear terms with linear-order ones. The resultant theory is equivalent to the standard perturbation
theory where we solve the perturbations up to the third order.
The advantage of using the closure
equations is that we can directly integrate the evolution equations for
the power spectrum numerically without using approximations
to derive solutions for perturbations. The analysis of the full
closure equations will be presented in a separate publication.

We solved the closure equations in DGP and $f(R)$ gravity
models as examples. In the DGP model, the two branches of the solutions were
considered. In the self-accelerating branch, the expansion of the
Universe is accelerated without having the cosmological constant.
In this branch, the BD parameter is negative and the linear growth
rate is suppressed compared with the dark energy model that follows
the same expansion history as the self-accelerating universe.
We found that the non-linear interactions in the Poisson equation
recover GR hence enhance the power spectrum
by shielding the BD scalar interactions. On the other hand, in
the normal branch solutions, where we need the cosmological
constant to accelerate the expansion of the universe, the BD
parameter is positive. Then the linear growth rate is enhanced
compared to the corresponding dark energy model. In this case,
the non-linearity of the Poisson equation
suppresses the power spectrum. In DGP models, it is also possible
to derive the semi-analytical solutions for the power spectrum
by using the Einstein-de Sitter approximations for the growth
rate. It was shown that the two methods give almost identical
results. In $f(R)$ gravity models, it is impossible to derive
semi-analytic solutions as the linear growth rate depends on
scales. In this case, we compare our results with N-body simulations.
Within the validity regime of the perturbation
theory inferred from an empirical formula derived in GR simulations,
our numerical solutions agree with N-body results very well.
In $f(R)$ gravity models, the BD parameter is vanishing and gravity
becomes strong below the Compton wavelength of the BD field and the
linear growth rate acquires a scale dependent enhancement.
Then the chameleon mechanism that recovers GR on small scales suppresses
the power spectrum.

Recently, the parametrization of the non-linear power spectrum in
modified gravity models was proposed within the framework of Parametrized
Post Friedman (PPF) formalism.
We checked whether the PPF formalism works or not
explicitly using the solutions in the perturbation theory.
In DGP models, we find that the PPF formalism works very well within
the validity regime of the perturbation theory by allowing a time
dependence in the PPF parameter $c_{\rm nl}$.
The power spectrum without the mechanism to recover
GR on small scales can be predicted by using a mapping formula between
the linear power spectrum and non-linear power spectrum derived in GR.
We used a fitting formula obtained by Smith et.al which
gives an accurate non-linear power spectrum in GR.
Using the value of $c_{\rm nl}$ obtained in the perturbation theory, we predicted
a fractional change in the non-linear power spectrum
in DGP relative to the equivalent dark energy models. With the
best fit cosmological parameters for SNe and CMB, there is a
$14\%$ change on linear scales and there is still a $7\%$ change
at $k=10h$Mpc$^{-1}$ at $z=0$. Interestingly, the maximum
change appears around $k=1h$Mpc$^{-1}$.
We applied the same procedure to $f(R)$ gravity models. It was found that
a slight modification was needed in the definition of $\Sigma(k,z)$
in the PPF formalism to reproduce the solutions in the perturbation theory. With this
modification, we could reproduce the perturbation theory solutions
very well again by allowing a time dependence in $c_{\rm nl}$.
Then we predicted the fully non-linear power spectrum
again using Smith et.al. fitting formula and the value of $c_{\rm nl}$
obtained in the perturbation theory.

We compared our PPF predictions with the N-body
results in both cases. It was found that the PPF formalism reproduces the
ratio between the power spectrum with and without the Vainshtein/chameleon
mechanism very well beyond the validity regime of the perturbation theory.
This indicates that the PPF formalism calibrated by the
perturbation theory describes the effect of the Vainshtein/chameleon
mechanism very well.
On the other hand, Smith et.al. fitting formula does not
predict the power spectrum without the Vainshtein/chameleon
mechanism very accurately. In $f(R)$ gravity,
this would be because the linear growth rate depends on
scales. An improvement of the prediction
can be made by obtaining the power spectrum without the Vainshtein/chameleon
mechanism more accurately. In order to demonstrate this fact,
we fitted the N-body results without the Vainshtein/chameleon mechanism
and applied the PPF formalism to predict the power spectrum
with the Vainsgtein/chameleon mechanism. In this way, it was shown
that the PPF formalism calibrated by the perturbation
theory reproduces the N-body results within the errors in
N-body simulations. It should be pointed out that N-body
simulations without the Veinshtein/chameleon mechanism is much easier
to perform as the modified gravity effects reduce to the
change of the Newton constant. On the other hand, if we need
to include the Veinshtein/chameleon mechanism properly, it is necessary
to solve the Klein-Gordon equation for the scalar field
directly.

The understanding of the non-linear power spectrum is essential
to distinguish between modified gravity models and dark energy
models. Especially, weak lensing is sensitive to the clustering
property on non-linear scales and all the predictions done by
neglecting the mechanism to recover GR on small scales should be revisited.
Our formalism based on the perturbation theory enables us to predict
the quasi non-linear power spectrum very accurately. The
PPF formalism calibrated by the perturbation theory gives an
analytical way to predict non-linear power spectrum. Of course, we
eventually need N-body simulations to check the predictions, but the
PPF formalism will provide a way to develop a fitting formula
for the non-linear power spectrum which is widely used to
predict weak lensing signal in GR dark energy models.
We tested our formalism in DGP and $f(R)$ gravity models
but we should bear in mind that these models face serious
difficulties. Our formalism is applicable to a wide
variety of modified gravity models and it is ready to use once
consistent models for modified gravity are developed.

\begin{acknowledgements}
We would like to thank Wayne Hu and Fabian Schmidt for
providing us their N-body data and for useful discussions.
KK is supported by ERC, RCUK and STFC. AT is supported
by a Grant-in-Aid for Scientific
Research from the Japan Society for the Promotion of
Science (JSPS) (No.~21740168). This work was supported in part by
Grant-in-Aid for Scientific Research on Priority Areas No.~467
``Probing the Dark Energy through an Extremely Wide and Deep Survey with
Subaru Telescope'', and JSPS Core-to-Core Program ``International
Research Network for Dark Energy''.
\end{acknowledgements}

\appendix
\section{Numerical algorithm}

In this paper, we follow the numerical scheme presented in
\cite{HT2009} to solve the closure equations derived in
\cite{Taruya:2007xy}. The basic procedure to solve
Eqs.~(\ref{eq:CLA_eq1})--(\ref{eq:CLA_eq3}) is the same as the one
summarised in Sec.III in \cite{HT2009}. In what follows, we
focus on the modifications that are necessary to apply it to
modified gravity models.

In \cite{HT2009}, the integration over $k$ has been performed
before performing the time evolution thanks to
the time-independent vertex functions which make
the integrands separable functions in terms of time and $k$.
In the present case, however, the integration over $k$ has to be performed
at each time step because the integrands are generally not separable
due to the time-dependent vertex functions. Another different
point from the previous work is the existence of the extra vertex
function, $\sigma_{2111}$.

Here we extend our numerical scheme presented in \cite{HT2009}
to include the above two new ingredients, the non-separable
integrands and the extra vertex function.
Firstly, we change the variable
$\bfk'$ to $\bfk-\bfk'$ in the integrations over $\bfk'$ in
Eqs.~(\ref{eq:CLA_eq1})--(\ref{eq:CLA_eq3}),
then we obtain a more symmetric form of the closure equations,
%
\begin{align}
\hat{\Lambda}_{ab}G_{bc}(k|\tau,\tau') &=
 \int^\tau_{\tau'}d\tau''\,M_{as}(k;\tau,\tau'')G_{sc}(k|\tau'',\tau')
  + S_{ar}(k;\tau)G_{rc}(k|\tau,\tau'),
  \label{eq:dG}\\
\hat{\Lambda}_{ab}R_{bc}(k;\tau,\tau') &=
 \int^\tau_{\tau_0}d\tau''\,M_{as}(k;\tau,\tau'')R_{\overline{sc}}(k;\tau'',\tau')
 +
 \int^{\tau'}_{\tau_0}d\tau''\,N_{a\ell}(k;\tau,\tau'')G_{c\ell}(k|\tau',\tau'')
  \notag \\
  & + S_{ar}(k;\tau)R_{rc}(k;\tau,\tau'), \label{eq:dR} \\
\hat{\Sigma}_{abcd}P_{cd}(k;\tau) &=
 \int^\tau_{\tau_0}d\tau''\,M_{as}(k;\tau,\tau'')R_{bs}(k;\tau,\tau'')
 +
 \int^\tau_{\tau_0}d\tau''\,N_{a\ell}(k;\tau,\tau'')G_{b\ell}(k|\tau,\tau'')
 \notag \\
&+ S_{ar}(k;\tau)P_{rb}(k;\tau)+ (a\leftrightarrow b),\label{eq:dP}
\end{align}
%
where
%
\begin{align}
 M_{as}(k;\tau,\tau'') &= 4\int\frac{d^3\kk'}{(2\pi)^3}
   \gamma_{apq}(\kk-\kk',\kk';\tau)\gamma_{\ell rs}(\kk'-\kk,\kk;\tau'')
   G_{q\ell}(k'|\tau,\tau'')R_{pr}(|\kk-\kk'|;\tau,\tau''),
       \label{eq:def_M}\\
 N_{a\ell}(k;\tau,\tau'') &= 2\int\frac{d^3\kk'}{(2\pi)^3}
   \gamma_{apq}(\kk-\kk',\kk';\tau)\gamma_{\ell rs}(\kk-\kk',\kk';\tau'')
   R_{qs}(k';\tau,\tau'')R_{pr}(|\kk-\kk'|;\tau,\tau''),
 \label{eq:def_N} \\
 S_{ar}(k;\tau) &= 3 \int\frac{d^3\bfk'}{(2\pi)^3}\sigma_{apqr}(\bfk',-\bfk',\bfk;\tau)P_{pq}(k';\tau), \label{eq:def_S}
\end{align}
%
and
%
\begin{equation}
 R_{\overline{sc}}(k;\tau'',\tau') \equiv
  \begin{cases}
    R_{sc}(k;\tau'',\tau'), & \tau''>\tau', \\
    R_{cs}(k;\tau',\tau''), & \tau''<\tau'.
  \end{cases}
\end{equation}
%
The domain of the two-dimensional integrand
(\ref{eq:def_M})--(\ref{eq:def_S}) is determined by
%
\begin{equation}
 k_{\rm min} \leq |\kk'|, |\kk-\kk'| \leq k_{\rm max}, \label{eq:GQlimit}
\end{equation}
%
where we set $k_{\rm min}=0.001h$Mpc$^{-1}$ and $k_{\rm max}=5h$Mpc$^{-1}$ according
to the convergence check performed in \cite{HT2009}.

\begin{figure}[h]
\centerline{
\includegraphics[width=8cm]{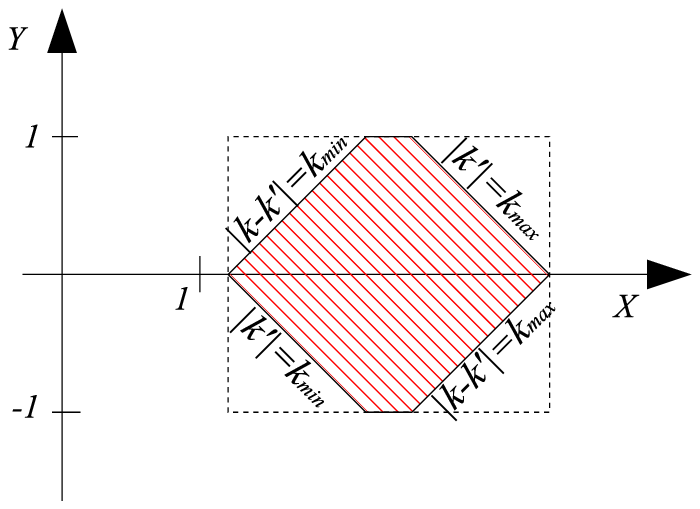}}
\caption{The domain of the integrations in the integrations
 (\ref{eq:def_M})--(\ref{eq:def_S}) in the $(X,Y)$-coordinate.
 The shadow region is defined by Eq.~(\ref{eq:GQlimit}) and, by
 definition, $-1\leq Y\leq 1$. The circumscribed rectangle is the computational
 domain for the Gaussian quadrature.}
\label{fig:GQ}
\end{figure}

Next we introduce $(X,Y,\phi)$-coordinates in the $\kk'$ space.
Suppose $\kk$ is aligned to the $k_z'$ axis in $\kk'$ space. Then
the elliptic coordinate is defined as
\begin{equation}
\kk' = \frac{\kk}{2} +\qq, \quad
\qq=\left(
\begin{array}{c}
\sinh \zeta \sin \mu \cos \phi \\
\sinh \zeta \sin \mu \sin \phi \\
\cosh \zeta \cos \mu
\end{array}
\right).
\end{equation}
We introduce the $XY$-coordinate as
\begin{equation}
X=\cosh \zeta, \quad Y= \cos \mu,
\end{equation}
where $X \geq 1$ and $-1 \leq Y \leq1$.
In these coordinates, the domain
(\ref{eq:GQlimit}) is depicted as the shaded hexagonal shape in
Fig.\ref{fig:GQ}. Then we perform the integration over the domain at each
time step. Note that the integration
over $\phi$ yields only a constant $2\pi$. In the previous work, we used the
trapezium rule for the integrations. Instead, in order to reduce the
computational cost without loosing the numerical accuracy, we implement the
two-dimensional Gaussian quadrature (GQ).
The two-dimensional GQ is adopted to
the circumscribed rectangle of the computational domain as shown in
Fig.~(\ref{fig:GQ}). The circumscribed rectangle is then transformed
to a unit square domain, $(x,y)\in[-1:1]\times[-1:1]$. On the unit domain,
the two-dimensional integration is approximated by
%
\begin{equation}
 \int^1_{-1}\int^1_{-1}f(x,y)\,dxdy \approx \sum^{N_{GQ}}_{i,j} f(x_i,y_i)w_iw_j,
\end{equation}
%
where $x_i$ and $y_i$ are defined as zero points of the Legendre
function, namely, $P_{N_{GQ}}(x_i)=0$. The weight $w_i$ is calculated by
%
\begin{equation}
 w_i = \frac{2}{(1-x_i^2)[P'_{N_{GQ}}(x_i)]^2},
\end{equation}
%
where the prime denotes derivative with respect to $x$ (e.g., see
\cite{NR}). The domain of the two-dimensional integrands (\ref{eq:GQlimit})
is extended to the unit domain so that the integrands vanish on the
extended region, and those on the grid $(x_i,y_j)$ are computed by
the cubic spline interpolation of the propagators and power
spectra. Having checked the convergence of the
numerical results, we fix the number of the grid points as $N_{GQ}=31$.

For the time evolutions, we set the initial redshift as $z_0=400$, and
the number of time steps as $N_z=400$, which are chosen so that the
final results become insensitive to these parameters at the order of
$0.1\%$. The number of wave numbers for the propagator and power spectra
is set to be $N_k=100$, and the discrete points in the wave number space
are taken so that they become dense around the baryon acoustic
oscillation scale.

\section{Perturbation theory and one-loop power spectra
in DGP model}

In this appendix, we derive a set of perturbative solutions up to the third order
in DGP models. Based on these solutions, we obtain the analytic expressions for
the one-loop power spectra.

\subsection{Solutions up to the third order}

The solutions for $\delta$ and $\theta$ are perturbatively expanded as
\begin{equation}
\delta(\bfk)=\delta^{(1)}(\bfk)+\delta^{(2)}(\bfk)+\delta^{(3)}(\bfk)+\cdots,
\quad
\theta(\bfk)=\theta^{(1)}(\bfk)+\theta^{(2)}(\bfk)+\theta^{(3)}(\bfk)+\cdots.
\end{equation}
In order to obtain the solutions analytically, we use the Einstein-de Sitter
(EdS) approximation (see sec.2.4.4 in \cite{Bernardeau:2001qr} for a review).
In the EdS approximation, all the non-linear growth rates
appearing in the higher-order solutions are approximately determined by the
linear growth rate $D_1(t)$. The explicit expressions for solutions at each
order can be obtained analytically, and are summarized below.
\bigskip

\noindent
\underline{1st-order solutions}:
\begin{equation}
\delta^{(1)}(\bfk;t)=D_1(t)\delta_0(\bfk),\quad
\theta^{(1)}(\bfk;t)=-(\dot{D}_1/H)\delta_0(\bfk),
\end{equation}
with $D_1$ being the linear growth rate. Here a dot denotes the derivative
with respect to time and $\delta_0(\bfk)$ denotes the primordial
density perturbation defined at early matter dominated era. We
assume $\delta_0(\bfk)$ obeys Gaussian statistic. The evolution equation for
$D_1$ is given by
\begin{equation}
\widehat{\mathcal{L}} D_1=0,
\end{equation}
where the linear operator $\widehat{\mathcal{L}}$ is given by
\begin{eqnarray}
\widehat{\mathcal{L}}\equiv\frac{d^2}{dt^2}+2H\frac{d}{dt}
-\frac{\kappa}{2}\rho_{m}\left(1+\frac{1}{3\beta}\right).
\end{eqnarray}

\noindent
\underline{2nd-order solutions}:
\begin{eqnarray}
\delta^{(2)}(\bfk;t)&=&\int\frac{d^3\bfk_1d^3\bfk_2}{(2\pi)^3}\,
\delta_{\rm D}(\bfk-\bfk_{12})\,\delta_0(\bfk_1)\delta_0(\bfk_2)
\Bigl[D_1^2(t)\,F^{(2)}_{\rm sym}(\bfk_1,\bfk_2)+
F_2(t)(1-\mu_{1,2}^2)\Bigr],
\\
\theta^{(2)}(\bfk;t)&=&\int\frac{d^3\bfk_1d^3\bfk_2}{(2\pi)^3}\,
\delta_{\rm D}(\bfk-\bfk_{12})\,\delta_0(\bfk_1)\delta_0(\bfk_2)\,
\Bigl[-(\dot{D_1}D_1/H)\,G^{(2)}_{\rm sym}(\bfk_1,\bfk_2)
-(\dot{F}_2/H)\,(1-\mu_{1,2}^2)\Bigr],
\end{eqnarray}
where we define
\begin{equation}
\mu_{i,j}\equiv\frac{\bfk_i\cdot\bfk_j}{|\bfk_i|\,|\bfk_j|}.
\end{equation}
The symmetrized kernels of the integrals,
$F^{(2)}_{\rm sym}$ and $G^{(2)}_{\rm sym}$, are
respectively defined as
\begin{eqnarray}
F^{(2)}_{\rm sym}(\bfk_1, \bfk_2) &=&
\frac{5}{14}\left\{ \alpha(\bfk_1,\bfk_2)+\alpha(\bfk_2,\bfk_1)\right\}+
\frac{1}{7}\beta(\bfk_1,\bfk_2),
\nonumber\\
G^{(2)}_{\rm sym}(\bfk_1, \bfk_2) &=&
\frac{3}{14}\left\{ \alpha(\bfk_1,\bfk_2)+\alpha(\bfk_2,\bfk_1)\right\}+
\frac{2}{7}\beta(\bfk_1,\bfk_2).
\nonumber
\end{eqnarray}
Note that in addition to the linear growth rate $D_1$, the second-order
solutions contain the new growth function $F_2$, which originates from
the non-linearity of the Poisson equation. The evolution equation for
$F_2$ is
\begin{equation}
\widehat{\mathcal{L}} F_2=
-\frac{r_{\rm c}^2}{6\beta^3}\left(\frac{\kappa\,\rho_{m}}{3}\right)^2\,
D_1^2.
\end{equation}

\noindent
\underline{3rd-order solutions}:
\begin{eqnarray}
\delta^{(3)}(\bfk;t)&=&\int\frac{d^3\bfk_1d^3\bfk_2d^3\bfk_3}{(2\pi)^6}\,
\delta_{\rm D}(\bfk-\bfk_{123})\,
\delta_0(\bfk_1)\delta_0(\bfk_2)\delta_0(\bfk_3)\,
\Bigl[D_1^3(t)\,F^{(3)}_{\rm sym}(\bfk_1,\bfk_2,\bfk_3)+
C_3(t)\,C_{\rm sym}(\bfk_1,\bfk_2,\bfk_3)
\Bigr.
\nonumber\\
&&\quad+\Bigl.
F_3(t)F_{\rm sym}(\bfk_1,\bfk_2,\bfk_3)+
I_3(t)I_{\rm sym}(\bfk_1,\bfk_2,\bfk_3)+
J_3(t)J_{\rm sym}(\bfk_1,\bfk_2,\bfk_3)+
K_3(t)K_{\rm sym}(\bfk_1,\bfk_2,\bfk_3)
\nonumber\\
&&\quad+\Bigl.
L_3(t)L_{\rm sym}(\bfk_1,\bfk_2,\bfk_3)\Bigr],
\label{eq:delta_3}
\\
\theta^{(3)}(\bfk;t)&=&\int\frac{d^3\bfk_1d^3\bfk_2}{(2\pi)^6}\,
\delta_{\rm D}(\bfk-\bfk_{123})\,
\delta_0(\bfk_1)\delta_0(\bfk_2)\delta_0(\bfk_3)\,
\Bigl[-(\dot{D}_1D_1^2/H)\,G_{\rm sym}^{(3)}(\bfk_1,\bfk_2,\bfk_3)
-(\dot{C}_3/H)\,C_{\rm sym}(\bfk_1,\bfk_2,\bfk_3)
\Bigr.
\nonumber\\
&&\quad-\Bigl.
\{(\dot{F}_3-\dot{D}_1F_2)/H\} F_{\rm sym}(\bfk_1,\bfk_2,\bfk_3) -
\{(\dot{I}_3-D_1\dot{F}_2)/H\} I_3(t)I_{\rm sym}(\bfk_1,\bfk_2,\bfk_3)
\nonumber\\
&&\quad-\Bigl.
(\dot{J}_3/H)J_{\rm sym}(\bfk_1,\bfk_2,\bfk_3)-
(\dot{K}_3/H)K_{\rm sym}(\bfk_1,\bfk_2,\bfk_3)-
(\dot{L}_3/H)L_{\rm sym}(\bfk_1,\bfk_2,\bfk_3)\Bigr].
\label{eq:theta_3}
\end{eqnarray}

The symmetrized kernels of integrals are
\begin{eqnarray}
F^{(3)}_{\rm sym}(\bfk_1,\bfk_2,\bfk_3)&=&\frac{1}{3}\left[
\frac{2}{63}\beta(\bfk_1,\bfk_{23})
\left\{\beta(\bfk_2,\bfk_3)+
\frac{3}{4}\Bigl(\alpha(\bfk_2,\bfk_3)+\alpha(\bfk_3,\bfk_2)\Bigr)\right\}
\right.
\nonumber\\
&&\quad\quad+\frac{1}{18}\alpha(\bfk_1,\bfk_{23})
\left\{\beta(\bfk_2,\bfk_3)+
\frac{5}{2}\Bigl(\alpha(\bfk_2,\bfk_3)+\alpha(\bfk_3,\bfk_2)\Bigr)\right\}
\nonumber\\
&&\quad\quad\quad\quad
\left.+\frac{1}{9}\alpha(\bfk_{23},\bfk_1)
\left\{\beta(\bfk_2,\bfk_3)+
\frac{3}{4}\Bigl(\alpha(\bfk_2,\bfk_3)+\alpha(\bfk_3,\bfk_2)\Bigr)\right\}
+\,\,\mbox{(cyclic perm.)}\right],
\nonumber\\
G^{(3)}_{\rm sym}(\bfk_1,\bfk_2,\bfk_3)&=&\frac{1}{3}\left[
\frac{2}{21}\beta(\bfk_1,\bfk_{23})
\left\{\beta(\bfk_2,\bfk_3)+
\frac{3}{4}\Bigl(\alpha(\bfk_2,\bfk_3)+\alpha(\bfk_3,\bfk_2)\Bigr)\right\}
\right.
\nonumber\\
&&\quad\quad+\frac{1}{42}\alpha(\bfk_1,\bfk_{23})
\left\{\beta(\bfk_2,\bfk_3)+
\frac{5}{2}\Bigl(\alpha(\bfk_2,\bfk_3)+\alpha(\bfk_3,\bfk_2)\Bigr)\right\}
\nonumber\\
&&\quad\quad\quad\quad
\left.+\frac{1}{21}\alpha(\bfk_{23},\bfk_1)
\left\{\beta(\bfk_2,\bfk_3)+
\frac{3}{4}\Bigl(\alpha(\bfk_2,\bfk_3)+\alpha(\bfk_3,\bfk_2)\Bigr)\right\}
+\,\,\mbox{(cyclic perm.)}\right],
\nonumber\\
C_{\rm sym}(\bfk_1,\bfk_2,\bfk_3)&=&\frac{1}{3}\left[
\beta(\bfk_1,\bfk_{23})\,(1-\mu_{2,3}^2)+\,\,\mbox{(cyclic perm.)}\right],
\nonumber\\
F_{\rm sym}(\bfk_1,\bfk_2,\bfk_3)&=&\frac{1}{3}\left[
\alpha(\bfk_1,\bfk_{23})\,(1-\mu_{2,3}^2)+\,\,\mbox{(cyclic perm.)}\right],
\nonumber\\
I_{\rm sym}(\bfk_1,\bfk_2,\bfk_3)&=&\frac{1}{3}\left[
\alpha(\bfk_{23},\bfk_1)\,(1-\mu_{2,3}^2)+\,\,\mbox{(cyclic perm.)}\right],
\nonumber\\
J_{\rm sym}(\bfk_1,\bfk_2,\bfk_3)&=&\frac{1}{3}\left[
(1-\mu_{1,23}^2)\beta(\bfk_2,\bfk_3)+\,\,\mbox{(cyclic perm.)}\right],
\nonumber\\
K_{\rm sym}(\bfk_1,\bfk_2,\bfk_3)&=&\frac{1}{3}\left[\frac{1}{2}
(1-\mu_{1,23}^2)\Bigl\{\alpha(\bfk_2,\bfk_3)+\alpha(\bfk_3,\bfk_3)\Bigr\}
+\,\,\mbox{(cyclic perm.)}\right],
\nonumber\\
L_{\rm sym}(\bfk_1,\bfk_2,\bfk_3)&=&\frac{1}{3}\left[
(1-\mu_{1,23}^2)(1-\mu^2_{2,3})
+\,\,\mbox{(cyclic perm.)}\right],
\nonumber
\end{eqnarray}

\bigskip
In the expressions (\ref{eq:delta_3}) and (\ref{eq:theta_3}),
new growth functions originating from non-linearity of the Poisson equation
appear. The evolution equations for these function are given by
\begin{eqnarray}
&&\widehat{\mathcal{L}}\,C_3=\dot{D}_1\dot{F}_2,
\nonumber\\
&&\widehat{\mathcal{L}}\,I_3=D_1(\ddot{F}_2+2H\dot{F}_2)+\dot{D}_1\dot{F}_2,
\nonumber\\
&&\widehat{\mathcal{L}}\,J_3=-\frac{r_{\rm c}^2}{6\beta^3}
\left(\frac{\kappa\rho_{m}}{3}\right)^2\frac{D_1^3}{7},
\nonumber\\
&&\widehat{\mathcal{L}}\,K_3=-\frac{r_{\rm c}^2}{6\beta^3}
\left(\frac{\kappa\rho_{m}}{3}\right)^2\frac{5D_1^3}{7},
\nonumber\\
&&\widehat{\mathcal{L}}\,L_3=-\frac{r_{\rm c}^2}{6\beta^3}
\left(\frac{\kappa\rho_{m}}{3}\right)^2D_1F_2+
\frac{r_{\rm c}^4}{9\beta^5}
\left(\frac{\kappa\rho_{m}}{3}\right)^3D_1^3.
\nonumber
\end{eqnarray}

\subsection{One-loop power spectra}

Using the perturbative solutions obtained above,
the power spectrum can be expressed as
\begin{equation}
P_{ab}(k;t)=P_{ab}^{(11)}(k;t) + P_{ab}^{(22)}(k;t) +P_{ab}^{(13)}(k;t)
+\cdots\quad\quad;\quad (a,b=\delta\,\, \mbox{or}\,\, \theta).
\end{equation}
The terms $P^{(11)}(k)$ imply the linear power spectrum, given by
\begin{equation}
P^{(11)}_{\delta\delta}(k;t) = D_1^2(t)\,P_0(k),\quad
P^{(11)}_{\delta\theta}(k;t) = -\frac{D_1(t)\dot{D}_1(t)}{H}\,P_0(k),\quad
P^{(11)}_{\theta\theta}(k;t) = \left(\frac{\dot{D}_1(t)}{H}\right)^2\,P_0(k),
\end{equation}
where the power spectrum $P_0(k)$ is defined by
\begin{equation}
\langle\delta_0(\bfk)\delta_0(\bfk')\rangle=
(2\pi)^3\delta_{\rm D}(\bfk+\bfk')\,P_0(|\bfk|).
\end{equation}
The terms $P^{(22)}(k)$ and $P^{(13)}(k)$ are the so-called one-loop
power spectra whose explicit expressions respectively become
\begin{eqnarray}
P^{(22)}_{\delta\delta}(k;t) &=&\frac{k^3}{(2\pi)^2}\int_0^{\infty}dx\,x^2
P_0(kx)\,\int_{-1}^{1}d\mu\,P_0\left(k\sqrt{1+x^2-2\mu x}\right)
\nonumber\\
&&\times 2\,
\Bigl[ D_1^4(t)\,\left\{\frac{3x+7\mu-10\mu^2x}{14x(1+x^2-2\mu x)}\right\}^2
+F_2^2(t)\,\left(\frac{\mu^2-1}{1+x^2-2\mu x}\right)^2\Bigr.
\nonumber\\
&&\quad\quad\Bigl.
+D_1^2(t)F_2(t)\,\frac{(3x+7\mu-10\mu^2x)(1-\mu^2)}{7x(1+x^2-2\mu x)^2}
\Bigr],
\end{eqnarray}
\begin{eqnarray}
P^{(22)}_{\delta\theta}(k;t) &=&\frac{k^3}{(2\pi)^2}\int_0^{\infty}dx\,x^2
P_0(kx)\,\int_{-1}^{1}d\mu\,P_0\left(k\sqrt{1+x^2-2\mu x}\right)
\nonumber\\
&&\times \,
\Bigl[ -\frac{\dot{D}_1D_1^3}{H}\,
\frac{(x-7\mu+6\mu^2x)(-3x-7\mu+10\mu^2x)}{98x^2(1+x^2-2\mu x)^2}
-\frac{\dot{F}_2F_2}{H}\,\frac{2(\mu^2-1)^2}{(1+x^2-2\mu x)^2}
\Bigr.
\nonumber\\
&&\Bigl.\quad\quad
-\frac{\dot{F}_2D_1^2}{H}\,
\frac{(-3x-7\mu+10\mu^2x)(\mu^2-1)}{7x(1+x^2-2\mu x)^2}
-\frac{\dot{D}_1D_1F_2}{H}\,
\frac{(x-7\mu+6\mu^2x)(\mu^2-1)}{7x(1+x^2-2\mu x)^2}
\Bigr],
\end{eqnarray}
\begin{eqnarray}
P^{(22)}_{\theta\theta}(k;t) &=&\frac{k^3}{(2\pi)^2}\int_0^{\infty}dx\,x^2
P_0(kx)\,\int_{-1}^{1}d\mu\,P_0\left(k\sqrt{1+x^2-2\mu x}\right)
\nonumber\\
&&\times 2\,
\Bigl[ \left(\frac{\dot{D}_1D_1}{H}\right)^2\,
\left\{\frac{(x-7\mu+6\mu^2x)}{14x(1+x^2-2\mu x)}\right\}^2
+\left(\frac{\dot{F}_2}{H}\right)^2\,\frac{(\mu^2-1)^2}{(1+x^2-2\mu x)^2}
\Bigr.
\nonumber\\
&&\Bigl.\quad\quad
+\frac{\dot{D}_1\dot{F}_2D_1}{H^2}\,
\frac{(x-7\mu+6\mu^2x)(\mu^2-1)}{7x(1+x^2-2\mu x)^2}
\Bigr],
\end{eqnarray}
for $P^{(22)}(k)$, and
\begin{eqnarray}
P^{(13)}_{\delta\delta}(k;t) &=&\frac{k^3}{(2\pi)^2}P_0(k)
\int_0^{\infty}dx\,P_0(kx)\,
\nonumber\\
&&\times\Bigl[
\,\,D_1^4(t)\,\frac{1}{252x^3}
\Bigl\{12x-158x^3+100x^5-42x^7+3(x^2-1)^3(7x^2+2)
\ln\left|\frac{1+x}{1-x}\right|
\Bigr\}\Bigr.
\nonumber\\
&&\quad\quad+D_1(t)C_3(t)\,\frac{1}{6x^3}\Bigl\{ 6x-16x^3-6x^5+3(x^2-1)^3
\ln\left|\frac{1+x}{1-x}\right|\Bigr\}
\nonumber\\
&&\quad\quad+D_1(t)I_3(t)\,\frac{1}{6x}\Bigl\{ 6x+16x^3-6x^5+3(x^2-1)^3
\ln\left|\frac{1+x}{1-x}\right|\Bigr\}
\nonumber\\
&&\quad\quad\Bigl.
+D_1(t)\left\{K_3(t)+L_3(t)\right\}\,
\frac{1}{12x^3}\Bigl\{ -6x+22x^3+22x^5-6x^7+
3(x^2-1)^4\ln\left|\frac{1+x}{1-x}\right|\Bigr\}
\Bigr],
\end{eqnarray} 
\begin{eqnarray}
P^{(13)}_{\delta\theta}(k;t) &=&\frac{k^3}{(2\pi)^2}P_0(k)
\int_0^{\infty}dx\,P_0(kx)\,
\nonumber\\
&&\times\Bigl[
\,\,-\frac{\dot{D}_1D_1^3}{H}\,\frac{1}{252x^3}
\Bigl\{24x-202x^3+56x^5-30x^7+3(x^2-1)^3(5x^2+4)
\ln\left|\frac{1+x}{1-x}\right|
\Bigr\}\Bigr.
\nonumber\\
&&\quad\quad-\frac{(D_1C_3)^\cdot}{H}\,
\frac{1}{12x^3}\Bigl\{ 6x-16x^3-6x^5+3(x^2-1)^3
\ln\left|\frac{1+x}{1-x}\right|\Bigr\}
\nonumber\\
&&\quad\quad-\frac{(D_1I_3)^\cdot-\dot{F}_2D_1^2}{H}\,
\frac{1}{12x}\Bigl\{ 6x+16x^3-6x^5+3(x^2-1)^3
\ln\left|\frac{1+x}{1-x}\right|\Bigr\}
\nonumber\\
&&\quad\quad\Bigl.
-\frac{\{D_1(K_3+L_3)\}^\cdot}{H}\,
\frac{1}{24x^3}\Bigl\{ -6x+22x^3+22x^5-6x^7+
3(x^2-1)^4\ln\left|\frac{1+x}{1-x}\right|\Bigr\}
\Bigr],
\end{eqnarray}
\begin{eqnarray}
P^{(13)}_{\theta\theta}(k;t) &=&\frac{k^3}{(2\pi)^2}P_0(k)
\int_0^{\infty}dx\,P_0(kx)\,
\nonumber\\
&&\times\Bigl[
\,\,\left(\frac{\dot{D}_1D_1}{H}\right)^2\,\frac{1}{84x^3}
\Bigl\{12x-82x^3+4x^5-6x^7+3(x^2-1)^3(x^2+2)
\ln\left|\frac{1+x}{1-x}\right|
\Bigr\}\Bigr.
\nonumber\\
&&\quad\quad
+\frac{\dot{D}_1\dot{C}_3}{H^2}\,
\frac{1}{6x^3}\Bigl\{ 6x-16x^3-6x^5+3(x^2-1)^3
\ln\left|\frac{1+x}{1-x}\right|\Bigr\}
\nonumber\\
&&\quad\quad
+\frac{\dot{D}_1(\dot{I}_3-\dot{F}_2D_1)}{H^2}\,
\frac{1}{6x}\Bigl\{ 6x+16x^3-6x^5+3(x^2-1)^3
\ln\left|\frac{1+x}{1-x}\right|\Bigr\}
\nonumber\\
&&\quad\quad\Bigl.
+\frac{\dot{D}_1(\dot{K}_3+\dot{L}_3)}{H^2}\,
\frac{1}{12x^3}\Bigl\{ -6x+22x^3+22x^5-6x^7+
3(x^2-1)^4\ln\left|\frac{1+x}{1-x}\right|\Bigr\}
\Bigr],
\end{eqnarray}
for $P^{(13)}(k)$.

The power spectrum without the non-linear interaction terms ${\cal I}$
can be obtained by putting $F_2=C_3=I_3=K_3=L_3=0$ in $P^{(22)}$ and
$P^{(13)}$.

\end{document}